\documentclass[useAMS,usenatbib]{mn2e}
\usepackage{epsfig}

\def\nodata{...}
\def\p1{\phantom{1}}

\def\simless{\mathbin{\lower 3pt\hbox
     {$\rlap{\raise 5pt\hbox{$\char'074$}}\mathchar"7218$}}}   %< or of order
\def\simmore{\mathbin{\lower 3pt\hbox
     {$\rlap{\raise 5pt\hbox{$\char'076$}}\mathchar"7218$}}}   %> or of order
\def\hide#1{}
\newcommand{\msun}{\ensuremath{{\rm M}_\odot}}
\title[Amplitude and coherence of the kHz QPOs]{On the maximum amplitude
and coherence of the kilohertz quasi-periodic oscillations in low-mass
X-ray binaries}

\author[M. M\'endez]{Mariano M\'endez$^{1}$\thanks{E-mail:
mariano@sron.nl}\\ 
$^{1}$SRON - Netherlands Institute for Space Research, Sorbonnelaan 2, 3584 
CA Utrecht, The Netherlands\\}

\begin{document}

\date{Accepted  Received ; in original form }

\pagerange{\pageref{firstpage}--\pageref{lastpage}} \pubyear{2005}

\maketitle

\label{firstpage}

\begin{abstract} I study the behaviour of the maximum rms fractional
amplitude, $r_{\rm max}$ and the maximum coherence, $Q_{\rm max}$, of
the kilohertz quasi-periodic oscillations (kHz QPOs) in a dozen low-mass
X-ray binaries. I find that: (i) The maximum rms amplitudes of the lower
and the upper kHz QPO, $r^{\ell}_{\rm max}$ and $r^{\rm u}_{\rm max}$,
respectively, decrease more or less exponentially with increasing
luminosity of the source; (ii) the maximum coherence of the lower kHz
QPO, $Q^{\ell}_{\rm max}$, first increases and then decreases
exponentially with luminosity, at a faster rate than both $r^{\ell}_{\rm
max}$ and $r^{\rm u}_{\rm max}$; (iii) the maximum coherence of the
upper kHz QPO, $Q^{\rm u}_{\rm max}$, is more or less independent of
luminosity; and (iv) $r_{\rm max}$ and $Q_{\rm max}$ show the opposite
behaviour with hardness of the source, consistent with the fact that
there is a general anticorrelation between luminosity and spectral
hardness in these sources. Both $r_{\rm max}$ and $Q_{\rm max}$ in the
sample of sources, and the rms amplitude and coherence of the kHz QPOs
in individual sources show a similar behaviour with hardness. This
similarity argues against the interpretation that the drop of coherence
and rms amplitude of the lower kHz QPO at high QPO frequencies in
individual sources is a signature of the innermost stable circular orbit
around a neutron star. I discuss possible interpretations of these
results in terms of the modulation mechanisms that may be responsible
for the observed variability.

\end{abstract}


%%%%%%%%%%%%%%%%%%%%%%%%%%%%%%%%%%%%%%%%%%%%%%%%%%%%%%%%%%%%%%%%%%%%%%%%%%%%%%%%%%%%%
%%%%%%%%%%%%%%%%%%%%%%%%%%%%%%%%%%%%%%%%%%%%%%%%%%%%%%%%%%%%%%%%%%%%%%%%%%%%%%%%%%%%%
%%
%% INTRODUCTION
%%
%%%%%%%%%%%%%%%%%%%%%%%%%%%%%%%%%%%%%%%%%%%%%%%%%%%%%%%%%%%%%%%%%%%%%%%%%%%%%%%%%%%%%
%%%%%%%%%%%%%%%%%%%%%%%%%%%%%%%%%%%%%%%%%%%%%%%%%%%%%%%%%%%%%%%%%%%%%%%%%%%%%%%%%%%%%

\section{Introduction}
\label{intro}

Kilohertz quasi-periodic oscillations (kHz QPOs) in the X-ray flux of
low-mass X-ray binaries have drawn much attention since their discovery,
about ten years ago. The reason for this continued interest is that
since they most likely reflect the motion of matter very close to the
neutron star (or black hole) primary in these systems, these QPOs may
provide one of the few direct ways of measuring effects that are unique
to the strong gravitational-field regime in these systems. In general
two simultaneous kHz QPOs are seen in the power density spectra of
low-mass X-ray binaries, the lower and the upper kHz QPO according to
how they appear sorted in frequency.

Most (but not all) of the work on QPOs in these years \citep[see][for a
review]{vanderklis-review} has focused on the frequencies of these QPOs,
because those frequencies provide insights into the dynamics of the
system. Several models have been proposed to explain the observed
relation between both kHz QPO frequencies \citep*{miller, lamb, stella1,
stella2, osherovich, titarchuk, abramowicz, rebusco}, their relation
with the spin of the neutron star \citep[e.g.,][]{miller, kluzniak} as
well as with other low-frequency variability \citep[e.g.,][]{stella1,
osherovich, titarchuk, titarchuk2, titarchuk3}, both low-frequency QPOs
and broad-band variability.

From the beginning there has been interest on the other two properties
of the kHz QPOs, their amplitude and coherence
\citep[e.g.,][]{vanderklis-scox-1, wijnands-cygx-2}; but systematic
studies of those other QPO properties started to take off slightly later
on \citep*{jonker-340+0, jonker-5-1, disalvo-1728, disalvo-1636,
vanstraaten-0614, vanstraaten-0614-1728, mendez-3srcs, vanstraaten-1608,
homan-17+2, kuznetsov-cygx-2, barret-1608, barret-1636, barret-nordita,
altamirano-1636}.

Recently, \cite{barret-1608, barret-1636, barret-nordita} carried out a
systematic study of the kHz QPO coherence and rms amplitude in three
X-ray binaries, 4U 1636--53, 4U 1608--52, and 4U 1735--44. They find
that in all three sources the coherence and rms amplitude of the lower
kHz QPO increase slowly with frequency, and after they reach their
maximum values they decrease abruptly as the QPO frequency keeps on
increasing. (The sudden drop is most noticeable in the coherence of the
lower kHz QPO; in the case of the rms amplitude the drop is less
abrupt.) They propose that this behaviour is due to effects related to
the innermost stable circular orbit, ISCO, around the neutron star in
these systems.

Triggered by these results, in this paper I investigate the dependence
of the maximum coherence and rms amplitude of both kHz QPOs in a large
sample of low-mass X-ray binaries. In \S\ref{data} I describe the data I
use in the rest of the paper. Since I collected most of the data from
the literature, I spend some time describing how those data were
obtained and the selection criteria. In \S\ref{results} I present the
results, and I discuss them in \S\ref{discussion}.  

%%%%%%%%%%%%%%%%%%%%%%%%%%%%%%%%%%%%%%%%%%%%%%%%%%%%%%%%%%%%%%%%%%%%%%%%%%%%%%%%%%%%%
%%%%%%%%%%%%%%%%%%%%%%%%%%%%%%%%%%%%%%%%%%%%%%%%%%%%%%%%%%%%%%%%%%%%%%%%%%%%%%%%%%%%%
%%
%% FIG. 1
%%
%%%%%%%%%%%%%%%%%%%%%%%%%%%%%%%%%%%%%%%%%%%%%%%%%%%%%%%%%%%%%%%%%%%%%%%%%%%%%%%%%%%%%
%%%%%%%%%%%%%%%%%%%%%%%%%%%%%%%%%%%%%%%%%%%%%%%%%%%%%%%%%%%%%%%%%%%%%%%%%%%%%%%%%%%%%

\begin{figure}
\centerline{\epsfig{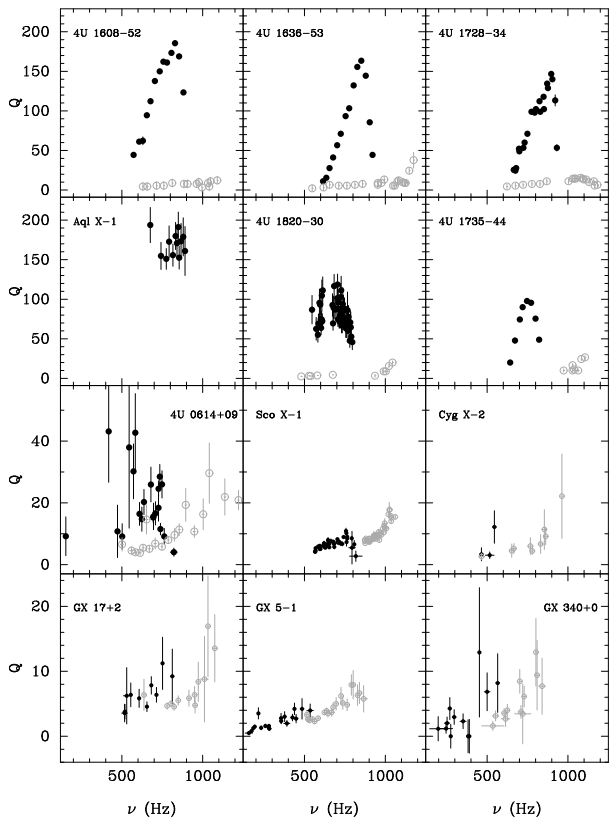}}     
\caption{The coherence of the lower (filled black symbols) and upper
(open gray symbols) kHz QPO as a function of the QPO frequency for the
sources studied in this paper. Notice that the range spanned by the
y-axis decreases from the panels at the top to those at the bottom.
\label{Q}}
\end{figure}

%%%%%%%%%%%%%%%%%%%%%%%%%%%%%%%%%%%%%%%%%%%%%%%%%%%%%%%%%%%%%%%%%%%%%%%%%%%%%%%%%%%%%
%%%%%%%%%%%%%%%%%%%%%%%%%%%%%%%%%%%%%%%%%%%%%%%%%%%%%%%%%%%%%%%%%%%%%%%%%%%%%%%%%%%%%
%%
%% DATA SELECTION AND ANALYSIS
%%
%%%%%%%%%%%%%%%%%%%%%%%%%%%%%%%%%%%%%%%%%%%%%%%%%%%%%%%%%%%%%%%%%%%%%%%%%%%%%%%%%%%%%
%%%%%%%%%%%%%%%%%%%%%%%%%%%%%%%%%%%%%%%%%%%%%%%%%%%%%%%%%%%%%%%%%%%%%%%%%%%%%%%%%%%%%

\section{Data selection and analysis}
\label{data}

\subsection{Data selection}
\label{selection}

All the data that I use in this paper were obtained over the last 10
years with the Proportional Counting Array, \cite*[PCA;][]{jahoda-pca},
on board the Rossi X-ray Timing Explorer \cite*[{\em
RXTE};][]{bradt-rxte}.

I collected most of these data from the literature. For this I searched,
for as many sources as possible, all published values of the rms
fractional amplitude and the coherence of the kHz QPOs. The coherence
$Q$ of a QPO is defined as $Q = \nu_{\rm QPO} / \lambda$, where $\nu_{\rm
QPO}$ and $\lambda$ are the frequency and the full-width at half-maximum
of the QPO. Some authors report $\lambda$ and $\nu$ instead of $Q$; in
those cases I calculate $Q$ using the previous formula. The rms
amplitude, $r$, is calculated from $P$, the integral from $0$ to
$\infty$ of the Fourier power under the QPO, and the source intensity,
$S$, as $r = 100 \times \sqrt{P/S}$ \citep{vanderklis-rms}; from this
definition, $r$ is expressed as a percent of the total intensity of the
source.

%%%%%%%%%%%%%%%%%%%%%%%%%%%%%%%%%%%%%%%%%%%%%%%%%%%%%%%%%%%%%%%%%%%%%%%%%%%%%%%%%%%%%
%%%%%%%%%%%%%%%%%%%%%%%%%%%%%%%%%%%%%%%%%%%%%%%%%%%%%%%%%%%%%%%%%%%%%%%%%%%%%%%%%%%%%
%%
%% FIG. 2
%%
%%%%%%%%%%%%%%%%%%%%%%%%%%%%%%%%%%%%%%%%%%%%%%%%%%%%%%%%%%%%%%%%%%%%%%%%%%%%%%%%%%%%%
%%%%%%%%%%%%%%%%%%%%%%%%%%%%%%%%%%%%%%%%%%%%%%%%%%%%%%%%%%%%%%%%%%%%%%%%%%%%%%%%%%%%%

\begin{figure}
\centerline{\epsfig{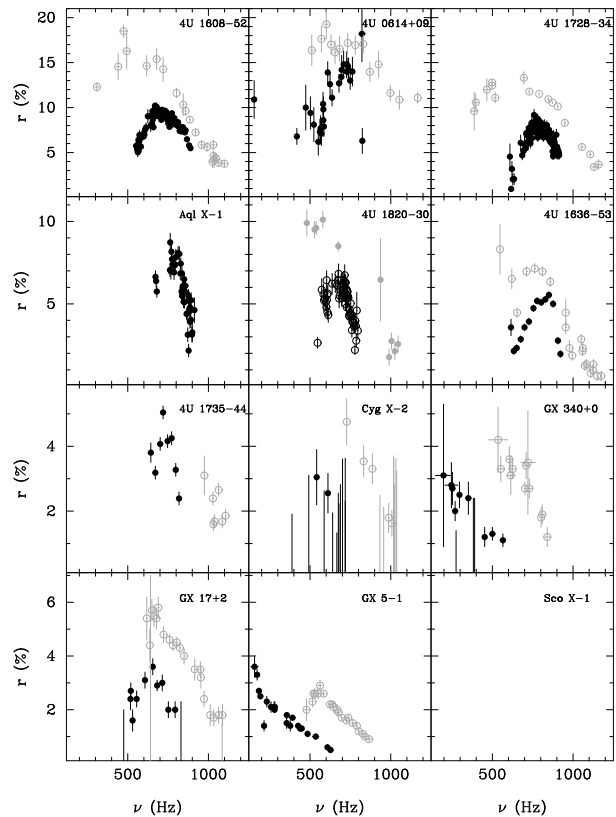}}
\caption{The rms amplitude of the lower (filled black symbols) and upper
(open gray symbols) kHz QPO as a function of the QPO frequency for the
sources studied in this paper. To avoid having to compress the scale of
the y-axis for some sources, the ordering of sources in this Figure is
different from that in Figure~\ref{Q}. There are no measurements of the
true rms amplitude of the QPOs in Sco X-1 (see \S~\ref{sources}), and
therefore I do not plot this source in this Figure.
\label{r}}
\end{figure}

%%%%%%%%%%%%%%%%%%%%%%%%%%%%%%%%%%%%%%%%%%%%%%%%%%%%%%%%%%%%%%%%%%%%%%%%%%%%%%%%%%%%%
%%%%%%%%%%%%%%%%%%%%%%%%%%%%%%%%%%%%%%%%%%%%%%%%%%%%%%%%%%%%%%%%%%%%%%%%%%%%%%%%%%%%%
%%
%% CONVENTION ON WHAT ARE KHZ QPOS
%%
%%%%%%%%%%%%%%%%%%%%%%%%%%%%%%%%%%%%%%%%%%%%%%%%%%%%%%%%%%%%%%%%%%%%%%%%%%%%%%%%%%%%%
%%%%%%%%%%%%%%%%%%%%%%%%%%%%%%%%%%%%%%%%%%%%%%%%%%%%%%%%%%%%%%%%%%%%%%%%%%%%%%%%%%%%%

From now on I will use the naming convention of the kHz QPOs introduced
by \cite*{belloni-2002}, in which the lower kHz QPO is called
$L_{\ell}$, and the upper kHz QPO is called $L_{\rm u}$. The frequency,
coherence and amplitude of each QPO carry a subscript or superscript
$\ell$ or $u$, respectively. A clarification is needed about what I call
kHz QPO in this paper. Some authors \cite[e.g.,][]{pbk} identify
low-frequency features as possibly being $L_{\ell}$ on the basis of an
extension of the QPO frequency-frequency correlation, $\nu_{\rm u} -
\nu_{\ell}$, to low frequencies (below $\nu_{\ell} \sim 50$ Hz). It
remains still uncertain whether these identifications are correct
\citep[cf][]{vanstraaten-1608, vanstraaten-1808}. In this paper I use
the expression ``kilohertz QPO'' to refer to features in the power
spectra of neutron star systems that have frequencies $> 150$ Hz, and
that have not been identified as hectohertz QPO
\cite[e.g.,][]{vanstraaten-0614}.

Since the goal is to compare the maximum values of $Q$ and $r$ of each
kHz QPO in different sources, one of the requirements in collecting the
data is that there are sufficient measurements of both parameters to
have confidence that the maximum values have been measured. This is in
part facilitated by the fact that we know the general dependence of $Q$
and $r$ on QPO frequency for both QPOs. For instance,
\cite{vanstraaten-0614, vanstraaten-0614-1728, vanstraaten-1608},
\cite{disalvo-1728, disalvo-1636}, \cite{mendez-3srcs},
\cite{barret-1608, barret-1636}, and \cite{altamirano-1636} show that
both $r_{\ell}$ and $Q_{\ell}$ increase with $\nu_{\ell}$, they peak at
intermediate to high values of $\nu_{\ell}$, and then they decrease as
$\nu_{\ell}$ increases \cite[see e.g., Figure 3 in][and Figure 3 in
\citealt{barret-1636}]{disalvo-1728}. On the other hand, $r_{\rm u}$ is
more or less flat at low values of $\nu_{\rm u}$ where it peaks, and
then decreases as $\nu_{\rm u}$ increases, while $Q_{\rm u}$ is more or
less constant with $\nu_{\rm u}$, although it increases towards higher
$\nu_{\rm u}$ values  \cite[see e.g., Figure 3 in ][and Figure 10 in
\citealt{altamirano-1636}]{disalvo-1728}. For this reason, here I
compile data from papers that present systematic studies of the QPO
properties, or that have enough measurements to deduce the systematic
trends of those properties. I do not use result from papers reporting
the discovery of kHz QPOs in a source, in which only one or two
measurements of the QPO properties are available. 

%%%%%%%%%%%%%%%%%%%%%%%%%%%%%%%%%%%%%%%%%%%%%%%%%%%%%%%%%%%%%%%%%%%%%%%%%%%%%%%%%%%%%
%%%%%%%%%%%%%%%%%%%%%%%%%%%%%%%%%%%%%%%%%%%%%%%%%%%%%%%%%%%%%%%%%%%%%%%%%%%%%%%%%%%%%
%%
%% TABLE L_{\ell}
%%
%%%%%%%%%%%%%%%%%%%%%%%%%%%%%%%%%%%%%%%%%%%%%%%%%%%%%%%%%%%%%%%%%%%%%%%%%%%%%%%%%%%%%
%%%%%%%%%%%%%%%%%%%%%%%%%%%%%%%%%%%%%%%%%%%%%%%%%%%%%%%%%%%%%%%%%%%%%%%%%%%%%%%%%%%%%

\begin{table*}
\centering
\begin{minipage}{145mm}
\caption{Maximum coherence and maximum rms amplitude of the lower kHz QPO for the
sources studied in this paper.}
\label{table_l}
\begin{tabular}{lcrccrcr}
\hline
Source      & $Q^{\ell}_{\rm max}$      & $\nu_\ell~$ & $(L/L_{\rm Edd})_{Q^{\ell}_{\rm max}}$ & $r^{\ell}_{\rm max}$   & $\nu_\ell~$ & $(L/L_{\rm Edd})_{r^{\ell}_{\rm max}}$ & Ref          \\
            &                           & (Hz)        &                                        & (\%)                   & (Hz)        &                                        &              \\
\hline
4U 0614+09  & $\p1   28.5 \pm \p1 4.0$  & 730         & $ 0.065 \pm 0.012                    $ & $\p1 9.3 \pm 0.3$      & 610         & $ 0.065 \pm 0.016                    $ &  1,2,3,4     \\
4U 1608--52 & $     247.0 \pm    16.0$  & 830         & $ 0.040 \pm 0.010                    $ & $   10.2 \pm 0.4$      & 670         & $ 0.030 \pm 0.075                    $ &  4,5,6,7,8   \\
4U 1636--53 & $     248.0 \pm    17.0$  & 840         & $ 0.085 \pm 0.021                    $ & $\p1 8.9 \pm 0.2$      & 780         & $ 0.085 \pm 0.021                    $ &  4,9,10,11   \\
4U 1728--34 & $     188.0 \pm    18.0$  & 900         & $ 0.080 \pm 0.020                    $ & $\p1 9.2 \pm 0.6$      & 760         & $ 0.070 \pm 0.017                    $ &  4,5,7,12    \\
4U 1735--44 & $     130.0 \pm    12.0$  & 770         & $ 0.120 \pm 0.030                    $ & $\p1 7.3 \pm 0.2$      & 740         & $ 0.120 \pm 0.030                    $ &  4,13,14     \\
4U 1820--30 & $     110.0 \pm    17.0$  & 700         & $ 0.200 \pm 0.050                    $ & $\p1 5.9 \pm 0.3$      & 600         & $ 0.200 \pm 0.050                    $ &  4,5         \\
Aql X-1     & $     190.0 \pm    19.0$  & 850         & $ 0.019 \pm 0.047                    $ & $\p1 8.7 \pm 0.6$      & 760         & $ 0.015 \pm 0.041                    $ &  4,5,7       \\
GX 5-1      & $\p1   13.7 \pm \p1 4.9$  & 620         & $ 1.800 \pm 0.450                    $ & $\p1 3.6 \pm 0.4$      & 160         & $ 1.100 \pm 0.255                    $ &  4,15        \\
GX 17+2     & $\p1   11.2 \pm \p1 4.0$  & 750         & $ 1.000 \pm 0.250                    $ & $\p1 3.6 \pm 0.3$      & 650         & $ 1.000 \pm 0.250                    $ &  4,16        \\
GX 340+0    & $\p1\p1 7.6 \pm \p1 2.4$  & 500         & $ 1.600 \pm 0.400                    $ & $\p1 2.8 \pm 0.8$      & 350         & $ 1.150 \pm 0.275                    $ &  4,17        \\
Cyg X-2     & \nodata                   & \nodata     & \nodata                                & $\p1 3.6 \pm 0.6$      & 520         & $ 0.800 \pm 0.200                    $ &  4,18        \\
%Cyg X-2     & $\p1   12.5 \pm \p1 5.3$  & 540         & $ 0.800 \pm 0.200                    $ & $\p1 3.6 \pm 0.6$      & 520         & $ 0.800 \pm 0.200                    $ &  4,18,19     \\
Sco X-1     & $\p1\p1 9.0 \pm \p1 0.2$  & 780         & $ 1.500 \pm 0.375                    $ & $>   1.4        $      & 590         & $ 1.500 \pm 0.375                    $ &  4,19        \\

\hline
\end{tabular}

NOTES ---
The $1$-$\sigma$ errors are given.  An aribtrary 25\% error is used for the luminosity (see text)

$Q^{\ell}_{\rm max}$ for 4U 0614+09 has been measured in the $4.6-60$ keV energy range.

The parameters for GX 340+0 and Cyg X-2 have been measured in the $5-60$ keV energy range.

The parameters for GX 17+2 have been measured in the $5.5-60$ keV energy range.

The parameters for Sco X-1 have been measured over a variable energy range (see text).

All the other parameters have been measured in the full PCA band.

REFERENCES ---
(1)  \cite{vanstraaten-0614};
(2)  \cite{vanstraaten-0614-1728};
(3)  \cite{ford-0614};
(4)  \cite{ford-et-al-2000};
(5)  {This paper};
(6)  \cite{barret-1608};
(7)  \cite{mendez-3srcs};
(8)  \cite{vanstraaten-1608};
(9)  \cite{disalvo-1636};
(10) \cite{barret-1636};
(11) \cite{barret-nordita};
(12) \cite{disalvo-1728};
(13) \cite{barret-nordita};
(14) \cite{ford-1735};
(15) \cite{jonker-5-1};
(16) \cite{homan-17+2};
(17) \cite{jonker-340+0};
%(18) \cite{kuznetsov-cygx-2};
(18) \cite{wijnands-cygx-2};
(19) \cite{vanderklis-scox-1}

\end{minipage}
\end{table*}

%%%%%%%%%%%%%%%%%%%%%%%%%%%%%%%%%%%%%%%%%%%%%%%%%%%%%%%%%%%%%%%%%%%%%%%%%%%%%%%%%%%%%
%%%%%%%%%%%%%%%%%%%%%%%%%%%%%%%%%%%%%%%%%%%%%%%%%%%%%%%%%%%%%%%%%%%%%%%%%%%%%%%%%%%%%
%%
%% POSSIBLE EDGE MAXIMA.
%%
%%%%%%%%%%%%%%%%%%%%%%%%%%%%%%%%%%%%%%%%%%%%%%%%%%%%%%%%%%%%%%%%%%%%%%%%%%%%%%%%%%%%%
%%%%%%%%%%%%%%%%%%%%%%%%%%%%%%%%%%%%%%%%%%%%%%%%%%%%%%%%%%%%%%%%%%%%%%%%%%%%%%%%%%%%%

In Figure~\ref{Q} and Figure~\ref{r} I present the plots of $Q$ and $r$
vs. QPO frequency for all the sources described in this paper: 4U
1608--52 \citep{barret-1608}, 4U 1636--52 \citep{barret-1636}, 4U
1728--34 \citep{barret-4srcs}, Aql X-1 \citep{mendez-3srcs} 4U 1820--30
\citep{barret-4srcs}, 4U 1735--34 \citep{barret-4srcs}, 4U 0614+09
\citep{vanstraaten-0614}, Sco X-1 \citep{vanderklis-scox-1}, Cyg X-2
\citep{wijnands-cygx-2, kuznetsov-cygx-2}, GX 17+2 \citep{homan-17+2},
GX 5-1 \citep{jonker-5-1}, and GX 340+0 \citep{jonker-340+0}. As it is
apparent from this figure, in most cases $Q_\ell$, $r_\ell$, and $r_{\rm
u}$ have well-defined maxima that occur at QPO frequencies in between
the minimum and maximum frequency observed for each QPO in each source.
For three sources (GX 17+2, GX 5-1, and GX 340+0), however, the maximum
value of $Q_\ell$ occurs at the edge of the frequency range. Similarly,
for two sources (GX 5-1 and GX 340+0), the maximum value of $r_\ell$,
and for three sources (4U 1820--30, Cyg X-2, and GX 340+0) the maximum
value of $r_{\rm u}$ occurs at the edge of the frequency range. For
$Q_{\rm u}$, it is more often the case that the maximum observed value
occurs at the edge of the frequency range (five out of eleven sources).
While it may still be possible that in these cases the maximum value of
$Q$ or $r$ has not yet been observed, on the basis of the slopes of the
relations of $Q$ and $r$ with $\nu$ in Figure~\ref{Q} and \ref{r}, it
seems very unlikely that the (unseen) maxima are too different from the
largest values so far measured. Except the case of $r_\ell$ in GX 5-1,
for which the maximum value might still be somewhat larger than the
largest value so far observed, for the other sources $Q$ and $r$ appear
to level off at the edge of the frequency range. Based on this, I
estimate that in those cases the maximum value can be at most $\sim
20$\% higher than the value I use in this paper; this potential
difference has no effect on the conclusions of the paper. Notice also
that since the significance at which a QPO is detected is proportional
to $r^2 \times Q^{1/2}$ \citep{vanderklis1989}, if $Q$ or $r$ were
significantly larger than so far observed, QPOs would have most likely
been detected above (below) the current maximum (minimum) frequencies.

%%%%%%%%%%%%%%%%%%%%%%%%%%%%%%%%%%%%%%%%%%%%%%%%%%%%%%%%%%%%%%%%%%%%%%%%%%%%%%%%%%%%%
%%%%%%%%%%%%%%%%%%%%%%%%%%%%%%%%%%%%%%%%%%%%%%%%%%%%%%%%%%%%%%%%%%%%%%%%%%%%%%%%%%%%%
%%
%% TABLE L_{\rm u}
%%
%%%%%%%%%%%%%%%%%%%%%%%%%%%%%%%%%%%%%%%%%%%%%%%%%%%%%%%%%%%%%%%%%%%%%%%%%%%%%%%%%%%%%
%%%%%%%%%%%%%%%%%%%%%%%%%%%%%%%%%%%%%%%%%%%%%%%%%%%%%%%%%%%%%%%%%%%%%%%%%%%%%%%%%%%%%

\begin{table*}
\centering
\begin{minipage}{145mm}
\caption{Maximum coherence and maximum rms amplitude of the upper kHz QPO for the
sources studied in this paper.}
\label{table_u}
\begin{tabular}{lcrccrcr}
\hline
Source      & $Q^{\rm u}_{\rm max}$ & $\nu_{\rm u}~$ & $(L/L_{\rm Edd})_{Q^{\rm u}_{\rm max}}$ & $r^{\rm u}_{\rm max}$  & $\nu_{\rm u}~$ & $(L/L_{\rm Edd})_{r^{\rm u}_{\rm max}}$ & Ref           \\
            &                       & (Hz)           &                                         & (\%)                   & (Hz)           &                                         &               \\
\hline
4U 0614+09  & $   10.9 \pm\p1 2.8$  & 1270           & $ 0.011 \pm 0.003                     $ & $   16.5 \pm 0.3$      &  510           & $ 0.005 \pm 0.001                     $ &  1,2,3,4      \\
4U 1608--52 & $   12.5 \pm\p1 2.8$  & 1060           & $ 0.050 \pm 0.012                     $ & $   18.5 \pm 0.7$      &  475           & $ 0.080 \pm 0.020                     $ &  4,5,6,7,8,9  \\
4U 1636--53 & $   52.0 \pm   14.0$  & 1230           & $ 0.120 \pm 0.030                     $ & $   17.1 \pm 0.6$      &  530           & $ 0.028 \pm 0.007                     $ &  4,5,10,11,12 \\
4U 1728--34 & $   10.5 \pm\p1 1.8$  & 1140           & $ 0.070 \pm 0.017                     $ & $   13.3 \pm 0.7$      &  600           & $ 0.027 \pm 0.007                     $ &  2,4,6,13     \\
4U 1735--44 &   \nodata             & \nodata        &   \nodata                               &    \nodata             & \nodata        &   \nodata                               &  \nodata      \\
4U 1820--30 & $   23.0 \pm\p1 5.0$  & 1050           & $ 0.250 \pm 0.062                     $ & $   10.1 \pm 0.3$      &  580           & $ 0.125 \pm 0.031                     $ &  4,5,14,15    \\
Aql X-1     &   \nodata             & \nodata        &   \nodata                               &    \nodata             & \nodata        &   \nodata                               &  \nodata      \\
GX 5-1      & $\p1 6.9 \pm\p1 0.8$  &  850           & $ 1.600 \pm 0.400                     $ & $\p1 2.9 \pm 0.2$      &  560           & $ 1.100 \pm 0.275                     $ &  4,16         \\
GX 17+2     & $   13.0 \pm\p1 3.5$  & 1010           & $ 1.000 \pm 0.250                     $ & $\p1 5.8 \pm 0.4$      &  590           & $ 0.900 \pm 0.225                     $ &  4,17         \\
GX 340+0    & $\p1 9.7 \pm\p1 2.8$  &  850           & $ 1.600 \pm 0.400                     $ & $\p1 4.2 \pm 1.0$      &  530           & $ 1.100 \pm 0.275                     $ &  4,18         \\
Cyg X-2     & $   22.2 \pm   13.6$  &  960           & $ 0.800 \pm 0.200                     $ & $\p1 3.5 \pm 0.4$      &  860           & $ 0.800 \pm 0.200                     $ &  4,19,20      \\
Sco X-1     & $   15.3 \pm\p1 0.3$  & 1040           & $ 1.500 \pm 0.375                     $ & $>   2.4        $      &  875           & $ 1.500 \pm 0.375                     $ &  4,21         \\

\hline
\end{tabular}

NOTES ---
The $1$-$\sigma$ errors are given. An aribtrary 25\% error is used for
the luminosity (see text)

The parameters for GX 340+0 and Cyg X-2 have been measured in the $5-60$
keV energy range.

The parameters for GX 17+2 have been measured in the $5.5-60$ keV energy
range.

The parameters for Sco X-1 have been measured over a variable energy
range (see text).

All the other parameters have been measured in the full PCA band.

REFERENCES ---
(1)  \cite{vanstraaten-0614};
(2)  \cite{vanstraaten-0614-1728};
(3)  \cite{ford-0614};
(4)  \cite{ford-et-al-2000};
(5)  {This paper};
(6)  \cite{mendez-3srcs};
(7)  \cite{barret-1608};
(8)  \cite{vanstraaten-1608};
(9)  \cite{gierlinski-1608};
(10) \cite{disalvo-1636};
(11) \cite{barret-1636};
(12) \cite{altamirano-1636}; 
(13) \cite{disalvo-1728};
(14) \cite{altamirano-1820};
(15) \cite{bloser-1820};
(16) \cite{jonker-5-1};
(17) \cite{homan-17+2};
(18) \cite{jonker-340+0};
(19) \cite{kuznetsov-cygx-2};
(20) \cite{wijnands-cygx-2};
(21) \cite{vanderklis-scox-1}

\end{minipage}
\end{table*}

%%%%%%%%%%%%%%%%%%%%%%%%%%%%%%%%%%%%%%%%%%%%%%%%%%%%%%%%%%%%%%%%%%%%%%%%%%%%%%%%%%%%%
%%%%%%%%%%%%%%%%%%%%%%%%%%%%%%%%%%%%%%%%%%%%%%%%%%%%%%%%%%%%%%%%%%%%%%%%%%%%%%%%%%%%%
%%
%% WAYS OF MEASURING Q/RMS AND PROBLEMS
%%
%%%%%%%%%%%%%%%%%%%%%%%%%%%%%%%%%%%%%%%%%%%%%%%%%%%%%%%%%%%%%%%%%%%%%%%%%%%%%%%%%%%%%
%%%%%%%%%%%%%%%%%%%%%%%%%%%%%%%%%%%%%%%%%%%%%%%%%%%%%%%%%%%%%%%%%%%%%%%%%%%%%%%%%%%%%

A word must be said about the different ways in which $Q$ and $r$ are
measured in these sources, and how these different ways of measuring
them can affect their values: 

\begin{enumerate}

\item The first QPO measurements were obtained over continuous {\em
RXTE} observations, usually stretches of up to $\sim 3000$-s exposure
time interrupted by the occultation of the source below the Earth's
horizon, or by the passage of the satellite by the South Atlantic
Anomaly, when many instruments were switched off for safety reasons.
Because the QPO frequency typically changes by a few tens of Hz within a
thousand seconds \citep{berger-1608}, the QPOs appear broader than they
actually are in the average power spectra of those observations. Since
the width measured in the average power spectrum, $\lambda_{\rm obs}$,
when a QPO of intrinsic width $\lambda$ drifts by an amount $\delta \nu$
is roughly $\lambda_{\rm obs} \approx \sqrt{\lambda^2 + (\delta
\nu)^2}$, clearly intrinsically narrower peaks are more affected by the
QPO frequency drift than intrinsically broader peaks. Because in general
$L_{\ell}$ is narrower than $L_{\rm u}$, the width of $L_{\ell}$ is the
most affected. On the other hand, since the total power in the QPO is
conserved in the averaging process, the rms amplitudes of the QPOs are
less affected by the QPO frequency drift. (Actually, the drift changes
the shape of the QPO in the average power spectrum, and because the
function used to fit them, usually a Lorentzian, no longer represents
them properly, also the rms amplitudes are affected, although to a
lesser degree).

\item When more data per source were collected, and it was realised 
that the QPO frequencies depended on other source parameters, it became
clear that a much better way of studying the QPO properties was to
average data taken at different epochs but having similar range of
values of those other source parameters. The typical examples are data
selection based on the colours of the source, or more specifically the
position along the track traced out by the source in a colour-colour or
colour-intensity diagram \cite[e.g.,][]{jonker-340+0, homan-17+2,
vanstraaten-0614, vanstraaten-0614-1728, disalvo-1728, disalvo-1636,
altamirano-1820, altamirano-1636}, or the frequency of one of the kHz
QPO, usually $L_{\ell}$ \cite[e.g.,][]{mendez-3srcs, barret-1636}, but
sometimes also $L_{\rm u}$ \citep{mendez-scox-1}, or the frequency of a
low-frequency QPO \cite[e.g.,][]{jonker-340+0}. While these selections
tend to reduce the spurious broadening of the QPOs due to the QPO
frequency drift, this effect may still be present in the results,
especially when the parameter used in selection spans a large interval. 

\end{enumerate}

In this paper I try to use as much as possible data collected according
to the second method just described, especially for the values of $Q$,
and in particular for $Q_{\ell}$, which would be the quantity that is
most affected by the averaging.

In a few cases one or more of the QPO parameters of a source are not
available in the literature; in some of those cases I myself have
systematically analysed an published other QPO parameters (e.g., QPO
frequency) of those sources, and since I still have the data available,
I analyse them myself for this paper to obtain those missing QPO
parameters. I describe those cases individually, including a summary of
the analysis procedures, below.

In Tables~\ref{table_l} and \ref{table_u} I provide all the measurements
of the maximum coherence, $Q_{\rm max}$, and maximum rms amplitude,
$r_{\rm max}$, for each QPO for each source, including the frequency of
the corresponding QPO at which $Q_{\rm max}$ and $r_{\rm max}$ occur; in
those Tables I also provide references to the papers from which these
values were collected.

\subsection{Measurements over different energy bands}
\label{energy}

%%%%%%%%%%%%%%%%%%%%%%%%%%%%%%%%%%%%%%%%%%%%%%%%%%%%%%%%%%%%%%%%%%%%%%%%%%%%%%%%%%%%%
%%%%%%%%%%%%%%%%%%%%%%%%%%%%%%%%%%%%%%%%%%%%%%%%%%%%%%%%%%%%%%%%%%%%%%%%%%%%%%%%%%%%%
%%
%% ENERGY BANDS USED AND EFFECTS
%%
%%%%%%%%%%%%%%%%%%%%%%%%%%%%%%%%%%%%%%%%%%%%%%%%%%%%%%%%%%%%%%%%%%%%%%%%%%%%%%%%%%%%%
%%%%%%%%%%%%%%%%%%%%%%%%%%%%%%%%%%%%%%%%%%%%%%%%%%%%%%%%%%%%%%%%%%%%%%%%%%%%%%%%%%%%%

In most cases in the literature, the QPO properties reported are those
measured over the full energy band of the PCA; nominally, this range is
$2 - 60$ keV. In any case, the PCA instrument is not very sensitive
above $\sim 30$ keV because of the lower effective area at those
energies, and because the background dominates the signal. In a few
cases (see Tables~\ref{table_l} and \ref{table_u}) measurements were
done over a different energy band to increase the sensitivity to kHz
QPOs; since the rms spectrum of the QPOs increases steeply with energy
\cite[see e.g.,][]{berger-1608}, in those other cases the authors
analysed the high-energy part of the data, above $\sim 4 - 5$ keV. On
the other hand, the ageing of the PCA units (PCU), and changes applied
to the gain voltage over the years to preserve the detectors affected
the energy scale of the PCA as well as its effective area as a function
of energy. The largest impact occurred at the low-energy end, since the
low-energy cut-off of the detectors is a strong function of the
instrument's high-voltage; from 1996 until present the lower energy
boundary of the PCA increased by $\sim 70\%$ due to the ageing of the
instrument and the gain changes applied to it\footnote{see
http://heasarc.gsfc.nasa.gov/docs/xte/e-c\_table.html}.

Because of the strong dependence of the rms spectrum of the QPOs on
energy, it is not straightforward to compare $r$ measurements over
different energy bands: The higher the low-energy boundary of the band
over which one measures $r$, the higher the value of $r$. On the other
hand, $Q$ measures the number of cycles of the oscillation which, at
least to first order, does not depend on energy. Therefore, one expects
that the values of $Q$ would not be much affected by the choice of the
energy band over which they are measured.

To confirm the above, and to assess the effect of the change of the
energy band on the parameters, I proceed as follows: I use the data for
4U 0614+09 from \cite{vanstraaten-0614-1728} who measured $Q$ and $r$ of
both kHz QPOs over the full PCA band ($2 - 60$ keV, nominally), and
\cite{vanstraaten-0614} who measured the same properties in the $4.6 -
60$ keV range. In both cases they report $Q$ and $r$ as a function of
QPO frequency, therefore I compare $Q$ and $r$ in both papers at the
same QPO frequency. I find that $Q$ is consistent with being the same in
both energy bands. On the other hand, the rms fractional amplitudes in
the $4.6 - 60$ keV band are $1.12 \pm 0.04$ ($1$-$\sigma$ error) times
higher than the rms fractional amplitudes in the full PCA band.

However, the energy spectrum of 4U 0614+09 is rather hard, and it may be
that in other sources that have softer spectra, the rms amplitudes in
different energy bands behave differently. To check this, I compiled rms
amplitudes in the $2-60$ band for 5 sources: 4U 1608--52
\citep{mendez-3srcs, vanstraaten-1608}, 4U 1728--34 \citep{mendez-1728,
vanstraaten-0614-1728}, 4U 1636--53 \citep{barret-1636}, 4U 0614+09
\citep{vanstraaten-0614}, and 4U 1820--30 \citep{altamirano-1820,
belloni-QPO-distrib}, and the corresponding rms amplitudes in the $5-60$
keV range \citep{jonker-0918}. I find that the rms fractional amplitude
in both bands are very well correlated (the correlation coefficient is
0.97), and that the rms amplitude in the $5-60$ keV band in these
sources is $1.25 \pm 0.02$ times larger than the rms amplitude in the
$2-60$ keV. This number is slightly higher, but still consistent (within
$3$-$\sigma$) with the value obtained above for 4U 0614+09.

In summary, the above results show that: (i) One can compare directly
$Q$-values in different energy bands, at least as long as the bands are
not too different from each other. Because of this, in the rest of the
paper I will compare $Q$ values even if they come from different bands;
(ii) changes of the low-end of the energy band affect the rms amplitudes
by $\sim 10 - 25\%$ of its value or less (in the case of instrument gain
changes and ageing, since changes of the low-energy boundary are smaller
than those in the case of 4U 0614+09 previously discussed, the effect is
probably much less important). In the next section, when I plot the QPO
parameters against luminosity and against each other, I divide the rms
amplitudes in the band going from $\sim 5$ to $60$ keV by $1.25$ to make
them compatible with the rms amplitude values obtained in the full PCA
band. The values quoted in Tables~\ref{table_l} and \ref{table_u},
however, are the ones actually measured and presented in the original
papers, in the energy band indicated there.

\subsection{Luminosity measurements}
\label{luminosity}

Once I have collected all $Q_{\rm max}$ and $r_{\rm max}$ values for
each QPO for each source, and the frequencies $\nu_{\ell}$ and $\nu_{\rm
u}$ at which those maximum values occur, I use Figure 1 in
\cite{ford-et-al-2000} to get the corresponding source luminosity: Using
the QPO frequency as input, I read off the luminosity of the source at
that frequency from that Figure. For three sources, 4U 1608--52, 4U
1636--53, and 4U 1820--30 (see below), the maximum rms amplitudes of
$L_{\rm u}$ occur in states of the source in which the QPO frequencies
are outside the range of frequencies in Figure 1 of
\cite{ford-et-al-2000}. In those three cases I either search the
literature for flux measurements of the source in that state, or I
extract spectra from the corresponding {\em RXTE} observations, and
calculate the luminosities myself, or both. Below I explain those three
cases in more detail.

%%%%%%%%%%%%%%%%%%%%%%%%%%%%%%%%%%%%%%%%%%%%%%%%%%%%%%%%%%%%%%%%%%%%%%%%%%%%%%%%%%%%%
%%%%%%%%%%%%%%%%%%%%%%%%%%%%%%%%%%%%%%%%%%%%%%%%%%%%%%%%%%%%%%%%%%%%%%%%%%%%%%%%%%%%%
%%
%% LUMINOSITY: UNCERTAINTIES
%%
%%%%%%%%%%%%%%%%%%%%%%%%%%%%%%%%%%%%%%%%%%%%%%%%%%%%%%%%%%%%%%%%%%%%%%%%%%%%%%%%%%%%%
%%%%%%%%%%%%%%%%%%%%%%%%%%%%%%%%%%%%%%%%%%%%%%%%%%%%%%%%%%%%%%%%%%%%%%%%%%%%%%%%%%%%%

The luminosities that I report here are uncertain for a number of
reasons: First, there is a statistical error in the fluxes derived from
model fitting. Given that these sources are quite bright when kHz QPOs
are detected, and that {\em RXTE} has a large effective area, these
errors are small, usually less than $5 - 10\%$. The second source of
error is the accuracy with which one can determine the luminosity using
the results of \cite{ford-et-al-2000} for a given QPO frequency. This is
not so much how accurately one can read $L$ from their plot, but the
fact that for each source there is not a single X-ray intensity or X-ray
flux value that corresponds to a given QPO frequency. This is the
so-called parallel-track phenomenon \cite[e.g.,][]{mendez-1608}. In the
extreme cases, the X-ray intensity of the source at a given QPO
frequency may by as much as a factor of 3 different \cite[e.g. 4U
1608--52,][]{mendez-1608}. On average, however, the range of intensities
at fixed QPO frequency is smaller, of the order of $20 - 50\%$
\citep{mendez-3srcs}. The third source of error is the use of the $2-50$
keV flux as a measure of the bolometric flux of the source. This is
discussed in detail by \cite{ford-et-al-2000}, and I therefore refer the
reader to that paper. This effect contributes uncertainties of the order
of $20-25\%$ of the reported luminosity values. The fourth source of
error, if $Q_{\rm max}$ and $r_{\rm max}$ depended not on the luminosity
of the source, $L$, but on the luminosity of the source normalized to
its own Eddington luminosity, $L/L_{\rm Edd}$, would be the use of a
single Eddington luminosity to normalise the observed luminosities of
all the sources in the sample, as done by \cite{ford-et-al-2000}. Of
course, if $Q_{\rm max}$ and $r_{\rm max}$ depended on $L$, and not on
$L/L_{\rm Edd}$, this would be of no concern. The value used by
\cite{ford-et-al-2000} is $L_{\rm Edd} = 2.5 \times 10^{38}
\mathrm{erg}~\mathrm{s}^{-1}$, which is the Eddington luminosity for a
$1.9 \msun$ neutron star accreting matter with cosmic abundance. Since
all these systems are thought to have had a common evolutionary path,
and to have accreted between $0.4$ and $0.8$ $\msun$, the use of a
single Eddington luminosity introduces an uncertainty that is of the
order of $25\%$ or less. (Notice that changes of the chemical
composition of the accreting material may introduce a difference of up
to a factor of 1.7 in the value of the Eddington luminosity; in the
unlikely case that the neutron stars in this sample had different
equations of state, the different redshifts at the surface of the
neutron star may add an extra factor 1.5.) Finally, the largest
uncertainty in the luminosity comes from the error in the estimate of
the distance to these sources \cite[see][for a discussion in the context
of these sources]{ford-et-al-2000}; this uncertainty can introduce
errors of up to 60\% in the luminosity \citep[see,
e.g.,][]{christian-swank}. It should be mentioned that for GX340+0 and
GX 5-1, \cite{christian-swank} only give upper limits to their
distances, hence for those two sources the luminosities in
\cite{ford-et-al-2000} should be taken as upper limits. 

In Tables \ref{table_l} and \ref{table_u} and in Figure \ref{low} I use
a fixed error of $25\%$ in the values of $L/L_{\rm Edd}$ as indicative
of the error in the luminosity. It is clear from the previous discussion
that this is a lower limit to the real error in this quantity.

\subsection{Individual sources}
\label{sources}

I now add a few remarks about special situations of some of the data for
sources I include in this paper:

%%%%%%%%%%%%%%%%%%%%%%%%%%%%%%%%%%%%%%%%%%%%%%%%%%%%%%%%%%%%%%%%%%%%%%%%%%%%%%%%%%%%%
%%%%%%%%%%%%%%%%%%%%%%%%%%%%%%%%%%%%%%%%%%%%%%%%%%%%%%%%%%%%%%%%%%%%%%%%%%%%%%%%%%%%%
%%
%% SPECIAL SITUATIONS ON INDIVIDUAL SOURCES
%%
%%%%%%%%%%%%%%%%%%%%%%%%%%%%%%%%%%%%%%%%%%%%%%%%%%%%%%%%%%%%%%%%%%%%%%%%%%%%%%%%%%%%%
%%%%%%%%%%%%%%%%%%%%%%%%%%%%%%%%%%%%%%%%%%%%%%%%%%%%%%%%%%%%%%%%%%%%%%%%%%%%%%%%%%%%%

\begin{enumerate}

%%%%%%%%%%%%%%%%%%%%%%%%%%%%%%%%%%%%%%%%%%%%%%%%%%%%%%%%%%%%%%%%%%%%%%%%%%%%%%%%%%%%%
%%%%%%%%%%%%%%%%%%%%%%%%%%%%%%%%%%%%%%%%%%%%%%%%%%%%%%%%%%%%%%%%%%%%%%%%%%%%%%%%%%%%%
%%
%% FIG. 3
%%
%%%%%%%%%%%%%%%%%%%%%%%%%%%%%%%%%%%%%%%%%%%%%%%%%%%%%%%%%%%%%%%%%%%%%%%%%%%%%%%%%%%%%
%%%%%%%%%%%%%%%%%%%%%%%%%%%%%%%%%%%%%%%%%%%%%%%%%%%%%%%%%%%%%%%%%%%%%%%%%%%%%%%%%%%%%

\begin{figure*}
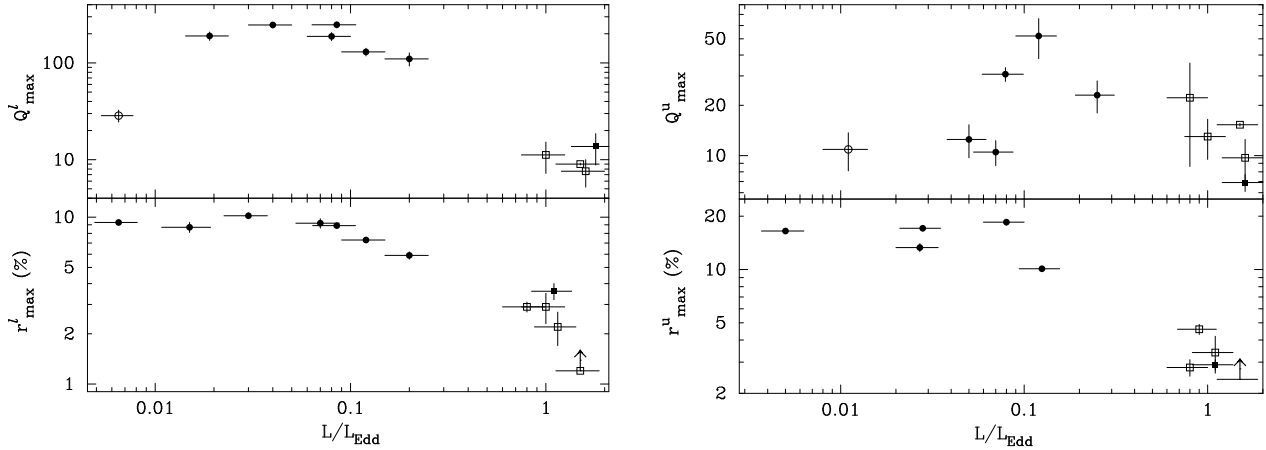

\centerline{\epsfig{file=low.ps,angle=-90,width=8.0cm}~~~~~~
            \epsfig{file=hig.ps,angle=-90,width=8.0cm}}	    
\caption{{\em Left panel}: The maximum coherence (upper panel) and
maximum rms amplitude (lower panel) of the lower kHz QPO for the sources
listed in Table~\ref{table_l}, as a function of the source luminosity at
the time at which those maximum values were reached. The luminosity is
in units of the Eddington luminosity for a $1.9 \msun$ neutron star.
Filled circles indicate measurements over the full energy band covered
by the PCA on board {\em RXTE}. Open circles indicate measurements over
a limited energy band (see Table~\ref{table_l} for details); in these
cases (except for Sco X-1), the rms amplitudes have been divided by 1.25
(see text for details). The rms amplitude in the case of Sco X-1 is not
corrected for dead-time, and hence is only a lower limit (indicated with
an arrow pointing upwards in the lower panel). {\em Right panel}: The
maximum coherence (upper panel) and maximum rms amplitude (lower panel)
of the upper kHz QPO for the sources listed in Table~\ref{table_u}, as a
function of the source luminosity at the time at which those maximum
values were reached. The luminosity is in units of the Eddington
luminosity for a $1.9 \msun$ neutron star. Symbols are the  same as in
the left panel. \label{low}}
\end{figure*}

%%%%%%%%%%%%%%%%%%%%%%%%%%%%%%%%%%%%%%%%%%%%%%%%%%%%%%%%%%%%%%%%%%%%%%%%%%%%%%%%%%%%%
%%%%%%%%%%%%%%%%%%%%%%%%%%%%%%%%%%%%%%%%%%%%%%%%%%%%%%%%%%%%%%%%%%%%%%%%%%%%%%%%%%%%%
%%
%% 4U 1608-52
%%
%%%%%%%%%%%%%%%%%%%%%%%%%%%%%%%%%%%%%%%%%%%%%%%%%%%%%%%%%%%%%%%%%%%%%%%%%%%%%%%%%%%%%
%%%%%%%%%%%%%%%%%%%%%%%%%%%%%%%%%%%%%%%%%%%%%%%%%%%%%%%%%%%%%%%%%%%%%%%%%%%%%%%%%%%%%

\item[{\bf 4U 1608--52}]: To measure $Q^{\ell}_{\rm max}$, I use the
same data in \cite{mendez-3srcs}. I accumulate power density spectra
every 64 seconds of data up to a Nyquist frequency of 2048 Hz; I average
a variable number of consecutive power spectra, in all cases less than
15, in order to obtain significant QPO detections, and I search for
high-frequency ($>$ 250 Hz) peaks. If I detect two QPO peaks, the one
with the lowest frequency is $L_{\ell}$; if I only detect one peak, I
decide whether it is $L_{\ell}$ or $L_{\rm u}$ on the basis of the the
rms-frequency relation \cite[e.g.,][]{mendez-3srcs}. If it is $L_{\rm
u}$ I discard it, if it is $L_{\ell}$ I measure $Q_{\ell}$. 

Individual $Q_{\ell}$ measurements have relatively large errors. Since
$Q_{\ell}$ is a function $\nu_{\ell}$, I then sort the results on the
basis of $\nu_{\ell}$, and I calculate average $Q_{\ell}$ values over a
narrow range of $\nu_{\ell}$. As usual, this procedure reduces the
errors in $Q_{\ell}$ at the expense of slightly underestimating its
maximum value.

I do not use the results of \cite{barret-1608} because in that case they
calculate $Q$ using whole observations; although they shift the
individual power spectra of each observation \citep[][see
below]{mendez-1608a}, still in many of those observations the frequency
of the QPO drifts by several tens of Hz, which has a negative effect in
the determination of $Q$. Nevertheless, I find a $Q^{\ell}_{\rm max}$
value that is larger than, but still consistent with, the value in
\cite{barret-1608}. 

For this source, $r^{\rm u}_{max}$ comes from \cite{vanstraaten-1608};
they find this rms amplitude at a QPO frequency $\nu_{\rm u}$ (see
Table~\ref{table_u}) that is outside the frequency range in
\cite{ford-et-al-2000}, and hence no simultaneous luminosity measurement
is available in \cite{ford-et-al-2000}. Van Straaten (2003) measured
this rms amplitude during their interval C; this corresponds to regions
1 and 2 in \cite{gierlinski-1608}, who find that the bolometric flux of
the source in those regions is $f_{\rm bol} = 1 \times 10^{-9}
\mathrm{erg~cm}^{-2}~\mathrm{s}^{-1}$. Interval C of
\cite{vanstraaten-1608} corresponds to the {\em RXTE} ObsId
30062-01-01-05. I therefore downloaded those data from the {\em RXTE}
archive, I produce a light curve of this observation and find that there
are no X-ray bursts or any anomaly in the data; I then extract an X-ray
spectrum using the Standard2 data following the procedures described on
the {\em RXTE} web
pages\footnote{http://heasarc.gsfc.nasa.gov/docs/xte}. I fit this
spectrum with a model consisting of a blackbody and a power law plus a
Gaussian emission line at around 6.5 keV, all affected by interstellar
absorption. The model fits the data well, with a $\chi^2$ per degree of
freedom of about 1. The $2-50$ keV unabsorbed flux of the source in this
observation is $1 \times 10^{-9} \mathrm{erg~cm}^{-2}~\mathrm{s}^{-1}$,
consistent with the flux of \cite{gierlinski-1608}. Using the same
distance to this source as in \cite{ford-et-al-2000}, $d=3.6$ Kpc, this
flux corresponds to a luminosity $L = 1.96 \times 10^{37} \mathrm{erg}~
\mathrm{s}^{-1}$, or $L/L_{\rm Edd} = 0.08$ using the Eddington
luminosity used by \cite{ford-et-al-2000}, $L_{\rm Edd} = 2.5 \times
10^{38} \mathrm{erg}~\mathrm{s}^{-1}$.

%%%%%%%%%%%%%%%%%%%%%%%%%%%%%%%%%%%%%%%%%%%%%%%%%%%%%%%%%%%%%%%%%%%%%%%%%%%%%%%%%%%%%
%%%%%%%%%%%%%%%%%%%%%%%%%%%%%%%%%%%%%%%%%%%%%%%%%%%%%%%%%%%%%%%%%%%%%%%%%%%%%%%%%%%%%
%%
%% 4U 1636-53
%%
%%%%%%%%%%%%%%%%%%%%%%%%%%%%%%%%%%%%%%%%%%%%%%%%%%%%%%%%%%%%%%%%%%%%%%%%%%%%%%%%%%%%%
%%%%%%%%%%%%%%%%%%%%%%%%%%%%%%%%%%%%%%%%%%%%%%%%%%%%%%%%%%%%%%%%%%%%%%%%%%%%%%%%%%%%%

\item[{\bf 4U 1636--53:}] To measure $Q^{\ell}_{\rm max}$, I use the
same data presented in \cite{disalvo-1636}. The procedure to produce
power spectra, identify the QPOs, and measure $Q_{\ell}$ is the same as
the one described for 4U 1608--52. The value of $Q^{\ell}_{\rm max}$
that I find is slightly higher than, but still consistent with, the one
found by \cite{barret-1636}.

For this source, $r^{\rm u}_{max}$ comes from \cite{altamirano-1636};
they find this rms amplitude at a QPO frequency $\nu_{\rm u}$ (see
Table~\ref{table_u}) that, as in the case of 4U 1608--52, is outside the
frequency range in \cite{ford-et-al-2000}, and hence no simultaneous
luminosity measurement is available in \cite{ford-et-al-2000}.
\cite{altamirano-1636} find this rms amplitude value during their
interval C, which corresponds to the {\em RXTE} ObsId 60032-05-10-000
and 90409-01-01-02. I produce a light curve of these observations and
find that there is an X-ray burst in the first one, and no X-ray bursts
or any anomaly in the data of the second one; I then follow the
procedures described on the {\em RXTE} web pages to extract an X-ray
spectrum from the Standard2 data of the second observation. I fit this
spectrum with a model consisting of a blackbody and a power law plus a
Gaussian emission line at around 6.5 keV, all affected by interstellar
absorption. The model fits the data well, with a $\chi^2$ per degree of
freedom of about 1. The $2-50$ keV unabsorbed flux of the source in this
observation is $1.9 \times 10^{-9}
\mathrm{erg~cm}^{-2}~\mathrm{s}^{-1}$. Using the same distance to this
source as in \cite{ford-et-al-2000}, $d=5.5$ Kpc, this flux corresponds
to a luminosity $L = 6.9 \times 10^{36} \mathrm{erg}~ \mathrm{s}^{-1}$,
or $L/L_{\rm Edd} = 0.028$ using the Eddington luminosity used by
\cite{ford-et-al-2000}, $L_{\rm Edd} = 2.5 \times 10^{38}
\mathrm{erg}~\mathrm{s}^{-1}$.

%%%%%%%%%%%%%%%%%%%%%%%%%%%%%%%%%%%%%%%%%%%%%%%%%%%%%%%%%%%%%%%%%%%%%%%%%%%%%%%%%%%%%
%%%%%%%%%%%%%%%%%%%%%%%%%%%%%%%%%%%%%%%%%%%%%%%%%%%%%%%%%%%%%%%%%%%%%%%%%%%%%%%%%%%%%
%%
%% 4U 1820-30
%%
%%%%%%%%%%%%%%%%%%%%%%%%%%%%%%%%%%%%%%%%%%%%%%%%%%%%%%%%%%%%%%%%%%%%%%%%%%%%%%%%%%%%%
%%%%%%%%%%%%%%%%%%%%%%%%%%%%%%%%%%%%%%%%%%%%%%%%%%%%%%%%%%%%%%%%%%%%%%%%%%%%%%%%%%%%%

\item[{\bf 4U 1820--30:}] To measure $Q^{\ell}_{\rm max}$ and $Q^{\rm
u}_{\rm max}$, I use the same data as in \cite{mendez-1820} and 
\cite{belloni-QPO-distrib}. The procedure to produce power spectra,
isolate $L_{\ell}$, and measure $Q_{\ell}$ is the same as the one
described for 4U 1608--52 and 4U 1636--53. To detect $L_{\rm u}$ and
measure $Q_{\rm u}$, I proceed as follows: I group the power spectra on
the basis of the frequency of $L_{\ell}$, $\nu_{\ell}$, such that within
each group $\nu_{\ell}$ does not change by more than 50 Hz. I then shift
the power spectra such that in each group the frequency of $L_{\ell}$ is
the same in all power spectra, and then I calculate an average power
spectrum per group. This procedure eliminates the effect of the drift in
$L_{\ell}$, and since the frequency difference in this source is more or
less constant when $\nu_{\ell}$ changes \citep{zhang-1820}, it also
corrects the effect of the drift in $L_{\rm u}$. I then fit the power
spectra with two Lorentzians that represent the QPOs, and a constant
that represents the Poisson counting noise. The fit yields, among other
parameters, $Q_{\rm u}$.

For this source, the $r^{\rm u}_{max}$ value is from
\cite{altamirano-1820}; they find this rms amplitude at a QPO frequency
$\nu_{\rm u}$ (see Table~\ref{table_u}) that, as in the case of 4U
1608--52 and 4U 1636--53, is outside the frequency range in
\cite{ford-et-al-2000}, and hence no simultaneous luminosity measurement
is available in \cite{ford-et-al-2000}. \cite{altamirano-1820} report
this rms amplitude at a position in the colour-colour diagram of 4U
1820--30 that is consistent with the interval $S_{a} = 1$ of
{\cite{bloser-1820}}. ($S_{a}$ is a parameter that measures the position
along the C-shaped track traced out by the source in a colour-colour
diagram.) During their observation, \cite{altamirano-1820} find that the
intensity of the source is 297 counts s$^{-1}$ PCU$^{-1}$, while in
their observation, \cite{bloser-1820} find the source at 312 counts
s$^{-1}$ PCU$^{-1}$. For their observation, \cite{bloser-1820} find a
$2-50$ keV luminosity of $2.25 -  2.31 \times 10^{37} \mathrm{erg}~
\mathrm{s}^{-1}$ for an assumed distance $d = 6.4$ Kpc.
\cite{ford-et-al-2000} use a distance to NGC 6624, the globular cluster
that contains 4U 1820--30, $d = 7.5$ Kpc. To be consistent with the
other luminosities for this source, which were taken from
\cite{ford-et-al-2000}, I convert the luminosity in \cite{bloser-1820}
to the one corresponding to the distance to 4U 1820--30 in
\cite{ford-et-al-2000}. This corresponds to $L = 3.15  \times 10^{37}
\mathrm{erg}~ \mathrm{s}^{-1}$, or $L/L_{\rm Edd} = 0.125$ using the
Eddington luminosity used by \cite{ford-et-al-2000}, $L_{\rm Edd} = 2.5
\times 10^{38} \mathrm{erg}~\mathrm{s}^{-1}$.

%%%%%%%%%%%%%%%%%%%%%%%%%%%%%%%%%%%%%%%%%%%%%%%%%%%%%%%%%%%%%%%%%%%%%%%%%%%%%%%%%%%%%
%%%%%%%%%%%%%%%%%%%%%%%%%%%%%%%%%%%%%%%%%%%%%%%%%%%%%%%%%%%%%%%%%%%%%%%%%%%%%%%%%%%%%
%%
%% FIG. 3
%%
%%%%%%%%%%%%%%%%%%%%%%%%%%%%%%%%%%%%%%%%%%%%%%%%%%%%%%%%%%%%%%%%%%%%%%%%%%%%%%%%%%%%%
%%%%%%%%%%%%%%%%%%%%%%%%%%%%%%%%%%%%%%%%%%%%%%%%%%%%%%%%%%%%%%%%%%%%%%%%%%%%%%%%%%%%%

\begin{figure*}
\centerline{\epsfig{file=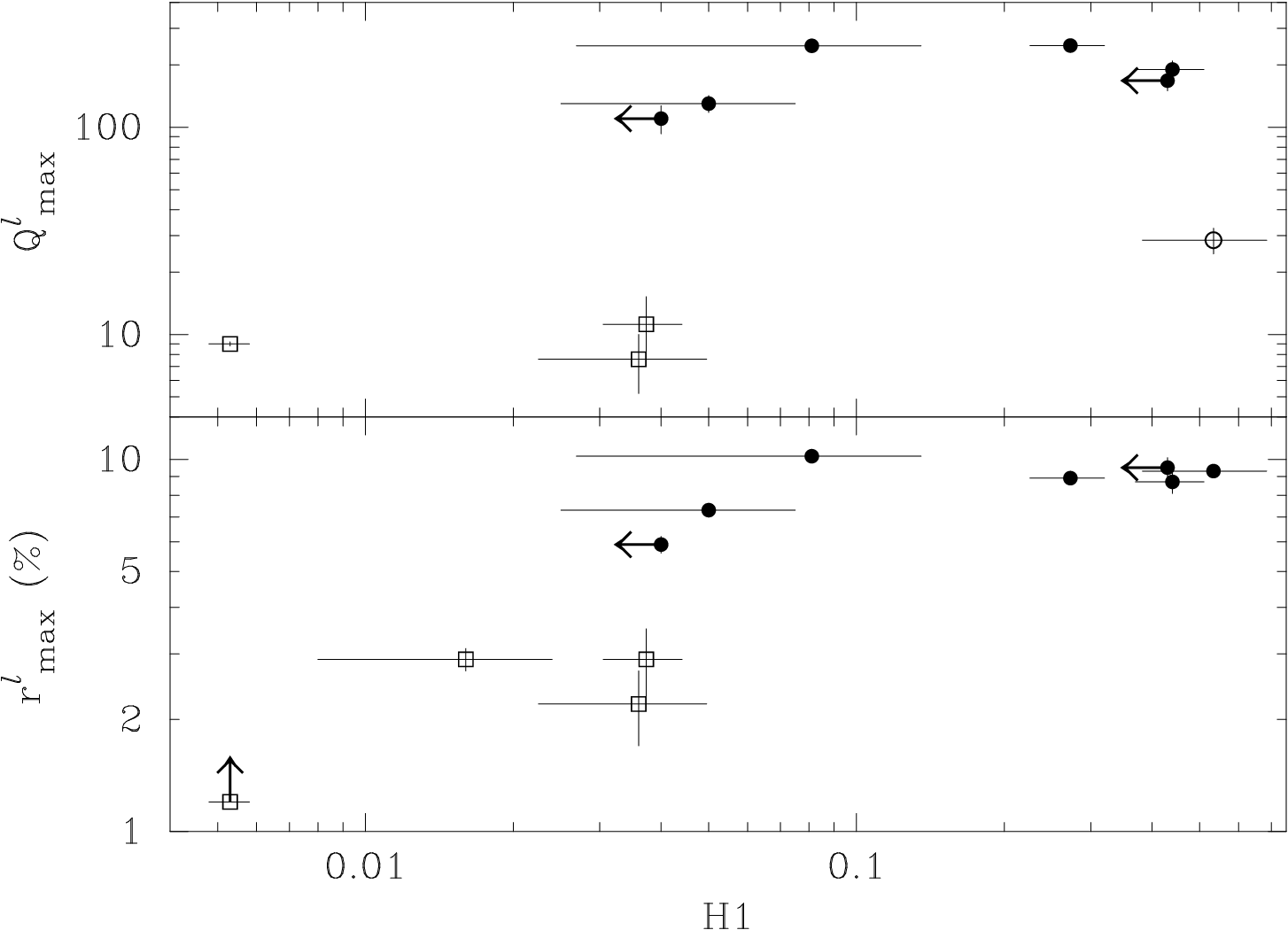,angle=-90,width=8.0cm}~~~~~~
            \epsfig{file=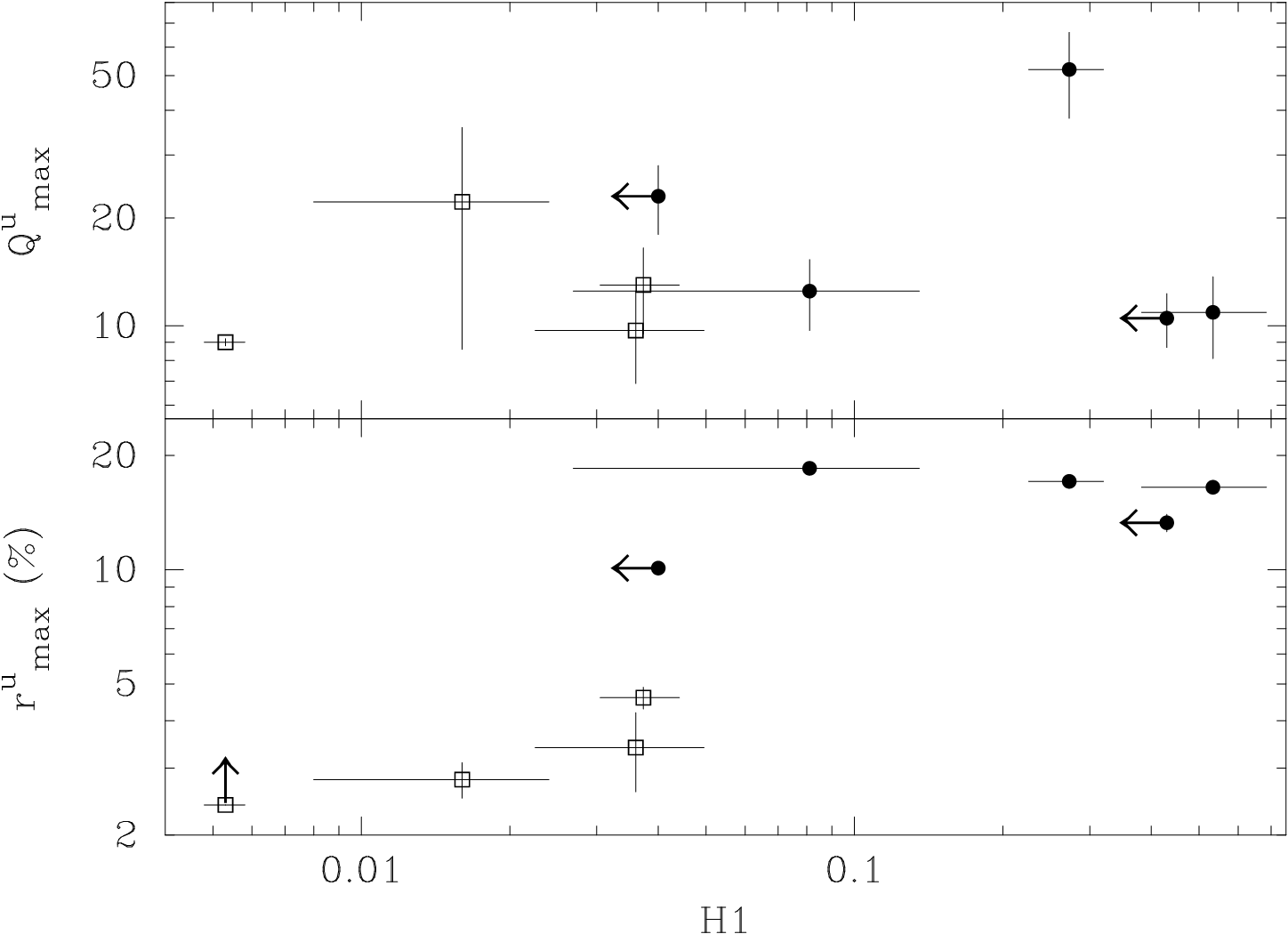,angle=-90,width=8.0cm}}
\caption{The maximum coherence (upper panels) and maximum rms amplitude
(lower panels) of the lower kHz QPO (left) and the upper kHz QPO (right)
for the sources listed in Tables~\ref{table_l} and \ref{table_u}, as a
function of the average source hardness, H1, defined as the ratio of the
$40-80$ keV to the $13-25$ keV count rate, measured with {\em HEAO-1}
\citep{levine-heao-1, vanparadijs-vanderklis}. For an explanation of the
symbols  see the caption of Figure~\ref{low}. The rms amplitudes in the
case of Sco X-1 are not corrected for dead-time, and hence are lower
limits (indicated with a vertical arrow in the lower panels). Upper
limits to the hardness are indicated with horizontal arrows pointing to
the left. I do not include GX 5-1 in this Figure because the {\em
HEAO-1} measurements of this source suffer contamination from a
previously unknown hard X-ray source in the field \citep[see explanation
in][] {levine-heao-1}, most likely GRS 1758--258 \citep*{gilfanov-1758}.
\label{h1}} 
\end{figure*}

%%%%%%%%%%%%%%%%%%%%%%%%%%%%%%%%%%%%%%%%%%%%%%%%%%%%%%%%%%%%%%%%%%%%%%%%%%%%%%%%%%%%%
%%%%%%%%%%%%%%%%%%%%%%%%%%%%%%%%%%%%%%%%%%%%%%%%%%%%%%%%%%%%%%%%%%%%%%%%%%%%%%%%%%%%%
%%
%% 4U 1728-34
%%
%%%%%%%%%%%%%%%%%%%%%%%%%%%%%%%%%%%%%%%%%%%%%%%%%%%%%%%%%%%%%%%%%%%%%%%%%%%%%%%%%%%%%
%%%%%%%%%%%%%%%%%%%%%%%%%%%%%%%%%%%%%%%%%%%%%%%%%%%%%%%%%%%%%%%%%%%%%%%%%%%%%%%%%%%%%

\item[{\bf 4U 1728--34:}] To measure $Q^{\ell}_{\rm max}$, I use the
same data presented in \cite{disalvo-1728} and \cite{mendez-3srcs}. The
procedure to produce power spectra, identify the QPOs, and measure
$Q_{\ell}$ is the same as the one described for 4U 1608--52. 

%%%%%%%%%%%%%%%%%%%%%%%%%%%%%%%%%%%%%%%%%%%%%%%%%%%%%%%%%%%%%%%%%%%%%%%%%%%%%%%%%%%%%
%%%%%%%%%%%%%%%%%%%%%%%%%%%%%%%%%%%%%%%%%%%%%%%%%%%%%%%%%%%%%%%%%%%%%%%%%%%%%%%%%%%%%
%%
%% 4U 1735-44
%%
%%%%%%%%%%%%%%%%%%%%%%%%%%%%%%%%%%%%%%%%%%%%%%%%%%%%%%%%%%%%%%%%%%%%%%%%%%%%%%%%%%%%%
%%%%%%%%%%%%%%%%%%%%%%%%%%%%%%%%%%%%%%%%%%%%%%%%%%%%%%%%%%%%%%%%%%%%%%%%%%%%%%%%%%%%%

\item[{\bf 4U 1735--44:}] For this source, \cite{barret-nordita}
recently reported a systematic study of $Q_{\ell}$ and $r_{\ell}$ as a
function of $\nu_{\ell}$; I therefore use $Q^{\ell}_{max}$ and
$r^{\ell}_{max}$ that are reported there. There are, however, very few
reports of the properties of $L_{\rm u}$ in the literature
\citep{wijnands-1735, ford-1735}; no reliable value of $Q_{\rm max}$ or
$r_{\rm max}$ are available (see also Fig. \ref{Q}). I therefore do not
discuss $L_{\rm u}$ of this source in this paper.

%%%%%%%%%%%%%%%%%%%%%%%%%%%%%%%%%%%%%%%%%%%%%%%%%%%%%%%%%%%%%%%%%%%%%%%%%%%%%%%%%%%%%
%%%%%%%%%%%%%%%%%%%%%%%%%%%%%%%%%%%%%%%%%%%%%%%%%%%%%%%%%%%%%%%%%%%%%%%%%%%%%%%%%%%%%
%%
%% Aql X-1
%%
%%%%%%%%%%%%%%%%%%%%%%%%%%%%%%%%%%%%%%%%%%%%%%%%%%%%%%%%%%%%%%%%%%%%%%%%%%%%%%%%%%%%%
%%%%%%%%%%%%%%%%%%%%%%%%%%%%%%%%%%%%%%%%%%%%%%%%%%%%%%%%%%%%%%%%%%%%%%%%%%%%%%%%%%%%%

\item[{\bf Aql X-1:}] For this source only one kHz QPO has been detected
\citep{zhang-aqlx-1, cui-aqlx-1}, which appears to be $L_{\ell}$
\citep{mendez-3srcs}, and hence there are no QPO parameters available
for $L_{\rm u}$.

%%%%%%%%%%%%%%%%%%%%%%%%%%%%%%%%%%%%%%%%%%%%%%%%%%%%%%%%%%%%%%%%%%%%%%%%%%%%%%%%%%%%%
%%%%%%%%%%%%%%%%%%%%%%%%%%%%%%%%%%%%%%%%%%%%%%%%%%%%%%%%%%%%%%%%%%%%%%%%%%%%%%%%%%%%%
%%
%% Cyg X-2
%%
%%%%%%%%%%%%%%%%%%%%%%%%%%%%%%%%%%%%%%%%%%%%%%%%%%%%%%%%%%%%%%%%%%%%%%%%%%%%%%%%%%%%%
%%%%%%%%%%%%%%%%%%%%%%%%%%%%%%%%%%%%%%%%%%%%%%%%%%%%%%%%%%%%%%%%%%%%%%%%%%%%%%%%%%%%%

\item[{\bf Cyg X-2:}] Since there are only three measurements of
$Q_\ell$ for Cyg X-2 (Fig.~\ref{Q}), I do not consider $Q_\ell$ for this
source in the rest of this paper. Also, there are only two measurements
of $r_\ell$ for this source, but the upper limits
\citep{wijnands-cygx-2} seem to indicate that the maximum $r_\ell$-value
has been measured, and hence here I use $r_\ell$ in the analysis.

%%%%%%%%%%%%%%%%%%%%%%%%%%%%%%%%%%%%%%%%%%%%%%%%%%%%%%%%%%%%%%%%%%%%%%%%%%%%%%%%%%%%%
%%%%%%%%%%%%%%%%%%%%%%%%%%%%%%%%%%%%%%%%%%%%%%%%%%%%%%%%%%%%%%%%%%%%%%%%%%%%%%%%%%%%%
%%
%% Sco X-1
%%
%%%%%%%%%%%%%%%%%%%%%%%%%%%%%%%%%%%%%%%%%%%%%%%%%%%%%%%%%%%%%%%%%%%%%%%%%%%%%%%%%%%%%
%%%%%%%%%%%%%%%%%%%%%%%%%%%%%%%%%%%%%%%%%%%%%%%%%%%%%%%%%%%%%%%%%%%%%%%%%%%%%%%%%%%%%

\item[{\bf Sco X-1:}] To avoid detector safety triggers, telemetry
saturation, and to reduce the dead-time effects produced by the high
count rate of Sco X-1, some observations of this source were carried out
with the source slightly off-axis, with some of the five proportional
counter units of the PCA switched off, recording only photons detected
by the upper anode chain of the PCA, recording only photons from a
limited energy range, or using a combination of these constraints
\citep{vanderklis-scox-1, mendez-scox-1}. Despite all these efforts, the
source count rate during these observations remained high, at a level in
which the dead-time of the PCA is still unknown. This means that the rms
values \cite[taken from][]{vanderklis-scox-1} are very uncertain, and in
fact could be larger than reported. For that reason, for Sco X-1 in this
paper I report the maximum rms amplitude as a lower limit.

\end{enumerate}

%%%%%%%%%%%%%%%%%%%%%%%%%%%%%%%%%%%%%%%%%%%%%%%%%%%%%%%%%%%%%%%%%%%%%%%%%%%%%%%%%%%%%
%%%%%%%%%%%%%%%%%%%%%%%%%%%%%%%%%%%%%%%%%%%%%%%%%%%%%%%%%%%%%%%%%%%%%%%%%%%%%%%%%%%%%
%%
%% RESULTS
%%
%%%%%%%%%%%%%%%%%%%%%%%%%%%%%%%%%%%%%%%%%%%%%%%%%%%%%%%%%%%%%%%%%%%%%%%%%%%%%%%%%%%%%
%%%%%%%%%%%%%%%%%%%%%%%%%%%%%%%%%%%%%%%%%%%%%%%%%%%%%%%%%%%%%%%%%%%%%%%%%%%%%%%%%%%%%

\section{Results:}
\label{results}

%%%%%%%%%%%%%%%%%%%%%%%%%%%%%%%%%%%%%%%%%%%%%%%%%%%%%%%%%%%%%%%%%%%%%%%%%%%%%%%%%%%%%
%%%%%%%%%%%%%%%%%%%%%%%%%%%%%%%%%%%%%%%%%%%%%%%%%%%%%%%%%%%%%%%%%%%%%%%%%%%%%%%%%%%%%
%%
%% REFERENCE TO FIG. 3
%%
%%%%%%%%%%%%%%%%%%%%%%%%%%%%%%%%%%%%%%%%%%%%%%%%%%%%%%%%%%%%%%%%%%%%%%%%%%%%%%%%%%%%%
%%%%%%%%%%%%%%%%%%%%%%%%%%%%%%%%%%%%%%%%%%%%%%%%%%%%%%%%%%%%%%%%%%%%%%%%%%%%%%%%%%%%%

Figure~\ref{low} shows the dependence of the maximum coherence and
maximum rms amplitude of both kHz QPOs as a function of source
luminosity in units of the Eddington luminosity for a $1.9 \msun$
neutron star. From this Figure it is apparent that $r^{\rm u}_{\rm max}$
and $r^{\ell}_{\rm max}$ both decrease more or less exponentially with
$L/L_{\rm Edd}$. For $r^{\ell}_{\rm max}$ the e-folding scale is
$L/L_{\rm Edd} = 0.86 \pm 0.06$, while $r^{\rm u}_{\rm max}$ decreases
significantly faster with $L/L_{\rm Edd}$ than $r^{\ell}_{\rm max}$,
with an e-folding scale $L/L_{\rm Edd} = 0.61 \pm 0.03$. Using roughly
the same sample of sources, \cite{jonker-0918} had already noticed that
the rms amplitude of the upper kHz QPO decreases as the luminosity of
the source increases.

From the same Figure, it is also apparent that at low luminosity,
$Q^{\ell}_{\rm max}$ first increases with $L$ up to $L/L_{\rm Edd} \sim
0.04$, and then it decreases exponentially, at a faster rate than
$r^{\ell}_{\rm max}$, with an e-folding scale $L/L_{\rm Edd} = 0.47 \pm
0.01$. On the other hand, $Q^{\rm u}_{\rm max}$ does not show any
significant trend with luminosity; although there is a hint that it
stays constant up to $L/L_{\rm Edd} \sim 0.07$, then it increases up to
$L/L_{\rm Edd} \sim 0.12$, and after that it decreases for increasing
luminosity, statistically a fit with a function that represents that
behavior is not significantly better than a fit of just a constant.

The gap in Figure~\ref{low} at $L/L_{\rm Edd} \sim 0.25 - 0.7$ separates
the Atoll sources at low $L$, and the Z sources at high $L$
\cite[see][for a definition of Atoll and Z sources]{hk89}. That gap
would be occupied by the four intermediate-type sources, GX 9+1, GX 9+9,
GX 3+1, and GX 13+1, which so far have not shown any kHz QPOs
\citep*{strohmayer-gx-sources, wijnands-gx-sources, homan-13+1,
schnerr-13+1, oosterbroek-3+1}. The upper limit to the rms amplitude of
the QPO in these sources ranges from 1.6\% to 2.6\%. Since the range of
luminosities spanned by these sources \citep{christian-swank} is
$L/L_{\rm Edd} \sim 0.12 - 0.44$, these upper limits are much lower than
would be expected from the interpolation of the trends of $r^{\ell}_{\rm
max}$ and $r^{\rm u}_{\rm max}$ with $L/L_{\rm Edd}$ in Figure
\ref{low}.

From Figure~\ref{low} it is apparent that the maximum rms amplitude of
both kHz QPOs and the maximum coherence of the lower kHz QPO are
consistently lower in the Z sources than in the Atoll sources. A valid
question would then be whether this difference could be due to a bias in
the way in which the QPOs of Z and Atoll sources are measured, combined
with the frequency drift of the QPOs (see \S\ref{selection}). A
comparison, however, of the width of the lower kHz QPO over a short time
interval in a Z and an Atoll source shows that this is not the case (as
I describe in \S2, the rms amplitudes are less affected by the frequency
drift). To see this, I compare an {\em RXTE} observation of the Z source
Sco X-1 from May 25 1996 \citep[see][]{vanderklis-scox-1,
mendez-scox-1}, and another {\em RXTE} observation of the atoll source
4U 1608--52 from March 26 1998 \citep[see][]{mendez-1608}. In the Sco
X-1 observation the power spectrum shows two simultaneous kHz QPOs that
move in frequency; $L_{\ell}$ moves from $\nu_{\ell} \sim 590$ Hz to
$\nu_{\ell} \sim 650$ Hz, and $L_{\rm u}$ moves from $\nu_{\rm u} \sim
890$ Hz to $\nu_{\rm u} \sim 950$ Hz. In the observation of 4U 1608--52
the power spectrum also shows two simultaneous kHz QPOs that move in
frequency. In this case $L_{\ell}$ moves from $\nu_{\ell} \sim 590$ Hz
to $\nu_{\ell} \sim 610$ Hz, and $L_{\rm u}$ moves from $\nu_{\rm u}
\sim 890$ Hz to $\nu_{\rm u} \sim 910$ Hz. To compare the QPOs in both
observations over the same frequency range, I calculate an average power
spectrum of the Sco X-1 data using only intervals in which $590
~\mathrm{Hz} \le \nu_{\ell} \le 610$ Hz; this corresponds to 24
individual power spectra each of them 16-s long. In the case of 4U
1608--52 I calculate an average power spectrum of the whole observation,
corresponding to 41 individual power spectra each of them 64-s long. To
correct for any residual drift in the QPO frequency, I apply the
shift-and-add procedure described in \cite{mendez-1608a}. The results,
however, do not change significantly if I average the individual power
spectra without first applying this method, most likely because the
frequency drift is small in both cases. From the above power spectra I
find that for Sco X-1 $Q_{\ell} = 4.2 \pm 0.4$, whereas for 4U 1608--52
$Q_{\ell} = 74.0 \pm 4.6$, which shows that in the Z source Sco X-1,
even over very short time intervals (in this case 384 s), the lower kHz
QPO is significantly broader than in the Atoll source 4U 1608--52. This
in turn shows that the drop of the maximum QPO coherence of the lower
kHz QPOs at high luminosities in Figure \ref{low} is real.

Table~\ref{table_l} shows that in all sources the maximum coherence
factor of the lower kHz QPO, $Q^{\ell}_{\rm max}$, occurs more or less
at the same frequency (the spread is $\sim 17\%$); Table~\ref{table_u}
shows that the same is true for the maximum coherence factor of the
upper kHz QPO, $Q^{\rm u}_{\rm max}$ (the spread in this case is $\sim
13\%$). Since the lifetime of an oscillation of frequency $\nu$ and
coherence factor $Q$ is roughly $\tau \sim Q / \nu$, the results
described above also indicate that the maximum lifetime of $L_{\ell}$,
$\tau^{\ell}_{\rm max}$, first increases and then decreases with
$L/L_{\rm Edd}$, whereas the maximum lifetime of $L_{\rm u}$, $\tau^{\rm
u}_{\rm max}$, is independent of $L/L_{\rm Edd}$. 

%%%%%%%%%%%%%%%%%%%%%%%%%%%%%%%%%%%%%%%%%%%%%%%%%%%%%%%%%%%%%%%%%%%%%%%%%%%%%%%%%%%%%
%%%%%%%%%%%%%%%%%%%%%%%%%%%%%%%%%%%%%%%%%%%%%%%%%%%%%%%%%%%%%%%%%%%%%%%%%%%%%%%%%%%%%
%%
%% FIG. 5
%%
%%%%%%%%%%%%%%%%%%%%%%%%%%%%%%%%%%%%%%%%%%%%%%%%%%%%%%%%%%%%%%%%%%%%%%%%%%%%%%%%%%%%%
%%%%%%%%%%%%%%%%%%%%%%%%%%%%%%%%%%%%%%%%%%%%%%%%%%%%%%%%%%%%%%%%%%%%%%%%%%%%%%%%%%%%%

\begin{figure}
\centerline{\epsfig{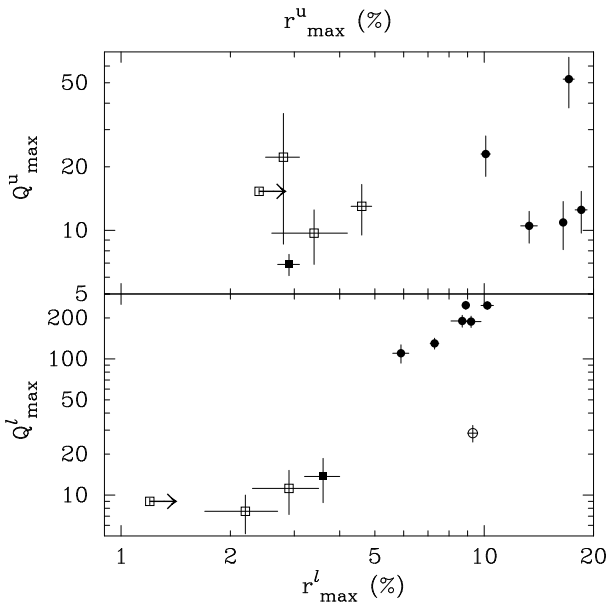}}
\caption{The maximum coherence of the kHz QPOs as a function of the 
maximum rms amplitude of the same kHz QPO, for the sources listed in
Tables~\ref{table_l} and \ref{table_u}; the upper and lower panel show
the  upper and the lower kHz QPO, respectively. For an explanation of
the symbols  see the caption of Figure~\ref{low}. In the case of Sco X-1
the rms amplitudes are not corrected for dead-time, and hence are lower
limits (indicated with horizontal arrows pointing to the right).
\label{rmsq}}
\end{figure}

%%%%%%%%%%%%%%%%%%%%%%%%%%%%%%%%%%%%%%%%%%%%%%%%%%%%%%%%%%%%%%%%%%%%%%%%%%%%%%%%%%%%%
%%%%%%%%%%%%%%%%%%%%%%%%%%%%%%%%%%%%%%%%%%%%%%%%%%%%%%%%%%%%%%%%%%%%%%%%%%%%%%%%%%%%%
%%
%% REFERENCE TO FIG. 4
%%
%%%%%%%%%%%%%%%%%%%%%%%%%%%%%%%%%%%%%%%%%%%%%%%%%%%%%%%%%%%%%%%%%%%%%%%%%%%%%%%%%%%%%
%%%%%%%%%%%%%%%%%%%%%%%%%%%%%%%%%%%%%%%%%%%%%%%%%%%%%%%%%%%%%%%%%%%%%%%%%%%%%%%%%%%%%

\cite*{vanparadijs-vanderklis} have shown that there is a general
correlation between the average source luminosity and the average source
hardness, H1, defined as the ratio of the count rate in the $40-80$ keV
band to the count rate in the $13-25$ keV band, measured with {\em
HEAO-1} \citep[see][]{levine-heao-1}. In that respect, as expected, the
plots of $Q_{\rm max}$ and $r_{\rm max}$ vs. hardness in Figure~\ref{h1}
show that $r^{\rm u}_{\rm max}$ and $r^{\ell}_{\rm max}$ both increase
with the hardness ratio H1, $Q^{\ell}_{\rm max}$ increases with H1 and
then it decreases for 4U 0614+09, the hardest source in this sample, and
as in the plots as a function of luminosity, $Q^{\rm u}_{\rm max}$ is
consistent with being constant with H1. At the risk of pointing out
something obvious, the opposite behaviour of the coherence and amplitude
of the QPOs to that in Figure~\ref{low} is due to the anticorrelation
between H1 and $L/L_{\rm Edd}$. I caution the reader that contrary to
the $L/L_{\rm Edd}$ values that I present in Tables \ref{table_l} and
\ref{table_u} and I plot in Figure~\ref{low}, for this Figure I use
average values of the spectral hardness measured several years before
the kHz QPOs were discovered. Notice that GX 340+0, for which the
distance, and hence the luminosity, is uncertain (see
\S~\ref{luminosity}), appears in Figure~\ref{h1} close to the other Z
sources (open symbols; GX 340+0 is the point at H1 $= 0.36$, just to the
left of GX 17+2 at H1 $= 0.37$); since H1 is a distance-independent
parameter, this suggests that the distance to GX 340+0 is not too much
in error.

%%%%%%%%%%%%%%%%%%%%%%%%%%%%%%%%%%%%%%%%%%%%%%%%%%%%%%%%%%%%%%%%%%%%%%%%%%%%%%%%%%%%%
%%%%%%%%%%%%%%%%%%%%%%%%%%%%%%%%%%%%%%%%%%%%%%%%%%%%%%%%%%%%%%%%%%%%%%%%%%%%%%%%%%%%%
%%
%% FIG. 6
%%
%%%%%%%%%%%%%%%%%%%%%%%%%%%%%%%%%%%%%%%%%%%%%%%%%%%%%%%%%%%%%%%%%%%%%%%%%%%%%%%%%%%%%
%%%%%%%%%%%%%%%%%%%%%%%%%%%%%%%%%%%%%%%%%%%%%%%%%%%%%%%%%%%%%%%%%%%%%%%%%%%%%%%%%%%%%

\begin{figure*}
\centerline{\epsfig{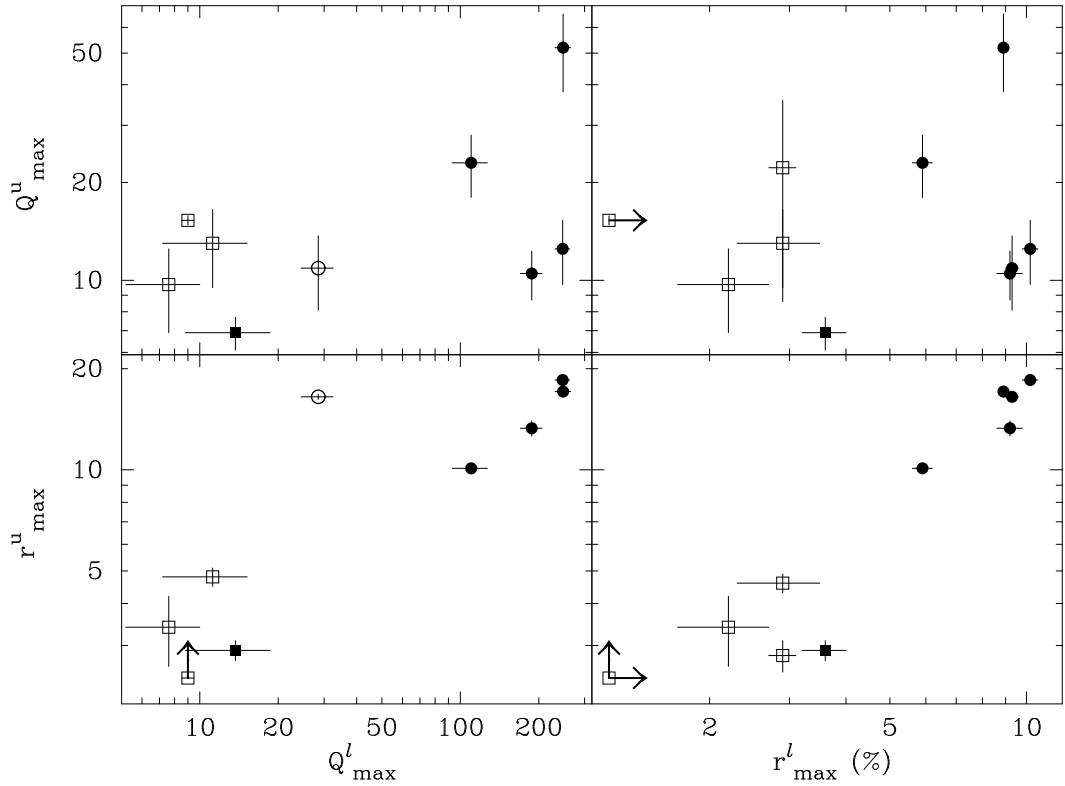}}
\caption{The maximum coherence and maximum rms amplitude of one kHz QPO
as a function of the maximum coherence and maximum rms amplitude of the
other kHz QPO, for the sources listed in Tables~\ref{table_l} and
\ref{table_u}.  The upper and lower left panels show plots,
respectively, of  $Q^{\rm u}_{\rm max}$ and $r^{\rm u}_{\rm max}$ vs.
$Q^{\ell}_{\rm max}$; the upper and lower right panels show plots,
respectively, of $Q^{\rm u}_{\rm max}$ and  $r^{\rm u}_{\rm max}$ vs.
$r^{\ell}_{\rm max}$. For an explanation of the symbols see the caption
of Figure~\ref{low}. In the case of Sco X-1 the rms amplitudes are not
corrected for dead-time, and hence are lower limits (indicated with
arrows pointing up and to the right). 
\label{cross-plots}} 
\end{figure*}

It is interesting to notice in this Figure that there is no gap in the
distribution of sources as a function of hardness between the Z and
Atoll sources. This is opposite to what is apparent in the plot of QPO
parameters vs. luminosity, where there is a gap corresponding to the
intermediate-type, the GX-sources (see above). The H1 values of the
GX-sources range from 0.03 to 0.14. However, both in Figure \ref{low}
and \ref{h1} there appears to be a gap in the values of $Q^{\ell}_{\rm
max}$, and perhaps also $r^{\ell}_{\rm max}$ and $r^{\rm u}_{\rm max}$
between the two types of sources (see also below).

%%%%%%%%%%%%%%%%%%%%%%%%%%%%%%%%%%%%%%%%%%%%%%%%%%%%%%%%%%%%%%%%%%%%%%%%%%%%%%%%%%%%%
%%%%%%%%%%%%%%%%%%%%%%%%%%%%%%%%%%%%%%%%%%%%%%%%%%%%%%%%%%%%%%%%%%%%%%%%%%%%%%%%%%%%%
%%
%% REFERENCE TO FIG. 5
%%
%%%%%%%%%%%%%%%%%%%%%%%%%%%%%%%%%%%%%%%%%%%%%%%%%%%%%%%%%%%%%%%%%%%%%%%%%%%%%%%%%%%%%
%%%%%%%%%%%%%%%%%%%%%%%%%%%%%%%%%%%%%%%%%%%%%%%%%%%%%%%%%%%%%%%%%%%%%%%%%%%%%%%%%%%%%

Figure~\ref{rmsq} shows the relation between $Q_{\rm max}$ and $r_{\rm
max}$ for the same QPO. To produce this Figure (and the next one), I
plot the $Q_{\rm max}$ vs. the $r_{\rm max}$ values for the same source,
even if they occur at slightly different luminosities. I also produced
plots of $Q_{\rm max}$ vs. $r_{\rm max}$ combining values for sources
that are closest in luminosity, even if this means combining values from
different sources, but the trends described below remain the same as
with the previous choice.

Except for the case of 4U 0614+09, $Q^{\ell}_{\rm max}$ and
$r^{\ell}_{\rm max}$ are positively correlated with each other.
Concerning 4U 0614+09, the hardest source in the sample and the one at
the lowest luminosity, it is as if in this case $Q^{\ell}_{\rm max}$
were too low for $r^{\ell}_{\rm max}$, or conversely, as if
$r^{\ell}_{\rm max}$ were too high for $Q^{\ell}_{\rm max}$. This is
already apparent in Figure~\ref{low}, where $L_{\ell}$ of 4U 0614+09
(low luminosity part of the plots) looses coherence without a similar
decrease of its rms amplitude. Figure~\ref{rmsq} also shows that $Q^{\rm
u}_{\rm max}$ is independent of $r^{\rm u}_{\rm max}$. 

Let us examine the case of the low $Q^{\ell}_{\rm max}$ value in 4U
0614+09 to check whether it is real: The measurements of $Q_{\ell}$ in
this source come from \cite{vanstraaten-0614}, and are made over a
different energy band than those of the other sources. For 4U 0614+09
\cite{vanstraaten-0614} use a band that starts at $4.6$ keV, whereas for
other sources at slightly higher luminosity the $Q_{\ell}$ values are
measured using the whole PCA band, nominally starting at $2$ keV (see
Table~\ref{table_l}). In \S\ref{energy} I compare measurements of
$Q_{\ell}$ as a function of $\nu_{\ell}$ (and also $Q_{\rm u}$ as a
function of $\nu_{\rm u}$) for this source over the full PCA band
\citep[$2-60$ keV;][]{vanstraaten-0614-1728} and over the $4.6-60$ keV
band \citep{vanstraaten-0614}, and I find that both sets of values are
compatible with each other. This indicates that most likely the
difference in $Q^{\ell}_{\rm max}$ between 4U 0614+09 and other sources
at slightly higher luminosity is real. Recently, Barret (private
communication) did a similar analysis on the data of 4U 0614+09 as he
did for instance for 4U 1636--53 \citep{barret-1636}, and he finds that
$Q_{\ell}$ in 4U 0614+09 is $\sim 6$ times smaller than in 4U 1636--53,
consistent with the results in Table~\ref{table_l}. This confirms that
$Q^{\ell}_{\rm max}$ in 4U 0614+09 is in fact smaller than in the other
sources at similar but slightly higher luminosity, and hence that the
decrease of $Q^{\ell}_{\rm max}$ at low luminosity is real.

%%%%%%%%%%%%%%%%%%%%%%%%%%%%%%%%%%%%%%%%%%%%%%%%%%%%%%%%%%%%%%%%%%%%%%%%%%%%%%%%%%%%%
%%%%%%%%%%%%%%%%%%%%%%%%%%%%%%%%%%%%%%%%%%%%%%%%%%%%%%%%%%%%%%%%%%%%%%%%%%%%%%%%%%%%%
%%
%% REFERENCE TO FIG. 6
%%
%%%%%%%%%%%%%%%%%%%%%%%%%%%%%%%%%%%%%%%%%%%%%%%%%%%%%%%%%%%%%%%%%%%%%%%%%%%%%%%%%%%%%
%%%%%%%%%%%%%%%%%%%%%%%%%%%%%%%%%%%%%%%%%%%%%%%%%%%%%%%%%%%%%%%%%%%%%%%%%%%%%%%%%%%%%

As I already mentioned, there appears to be a gap between the
$Q^{\ell}_{\rm max}$ and $r^{\ell}_{\rm max}$ values of the Z and Atoll
sources, with the coherence showing the largest gap (Figure 3, lower
panel). 4U 0614+09 appears to be the only source to (partially) break
this rule, since it has a $Q^{\ell}_{\rm max}$ value that is
intermediate between those of Z and Atoll sources. Similarly, the upper
panel of Figure 3 shows a gap between the $r^{\rm u}_{\rm max}$ values,
but not for the $Q^{\rm u}_{\rm max}$ values, of the Z and Atoll
sources. While this can indicate a dependence on luminosity (see
Fig.~\ref{low}) or on spectral hardness (see Fig.~\ref{h1}), this could
also point to a difference between Z and atoll sources. It is worth
noting, however, that there is still a trend of $Q^\ell_{\rm max}$ and
both rms amplitudes {\em within} the atoll sources in Figure~\ref{low}.
Furthermore, there is a significant trend in the relations between
$Q_{\rm max}$ and $r_{\rm max}$ within the atoll sources; e.g., the
relation between $Q^\ell_{\rm max}$ and $r^\ell_{\rm max}$ (lower panel
of Figure~\ref{rmsq}) is $8\sigma$ different from a constant. All this
suggests that the distinction between Z and Atoll sources cannot be the
(only) explanation for this difference.

The plots of $Q_{\rm max}$ and $r_{\rm max}$ of one QPO vs. the same
parameters of the other QPO are shown in Figure~\ref{cross-plots}. From
this Figure there are three things apparent: (i) $Q^{\rm u}_{\rm max}$
is independent both of $Q^{\ell}_{\rm max}$ and $r^{\ell}_{\rm max}$
(upper panels); (ii) as in the case of the plot of $r^{\ell}_{\rm max}$
vs. $Q^{\ell}_{\rm max}$, except for 4U 0614+09, $r^{\rm u}_{\rm max}$
is positively correlated with $Q^{\ell}_{\rm max}$. As in the case of
$L_{\ell}$, it appears as in 4U 0614+09, the source with the lowest
luminosity in the sample, the amplitude of $L_{\rm u}$, $r^{\rm u}_{\rm
max}$, were too high for the coherence of $L_{\ell}$, $Q^{\ell}_{\rm
max}$, or alternatively, the coherence of $L_{\ell}$, $Q^{\ell}_{\rm
max}$, were too low for the amplitude of $L_{\rm u}$, $r^{\rm u}_{\rm
max}$ (lower left panel). Finally, (iii) the amplitudes of both kHz QPOs
are positively correlated with each other, including in this case the
low-luminosity source 4U 0614+09 (lower right panel).

From the lower panels of this Figure, it is also apparent that there is
a gap between the $r^{\rm u}_{\rm max}$ values of the Z and Atoll
sources, similar to the ones described above

To summarize the results from Figures \ref{rmsq} and \ref{cross-plots},
the maximum amplitude and coherence of $L_{\ell}$, $r^{\ell}_{\rm max}$
and $Q^{\ell}_{\rm max}$, and the maximum rms amplitude of $L_{\rm u}$,
$r^{\rm u}_{\rm max}$, are all correlated with each other (in the case
of $Q^{\ell}_{\rm max}$, at least above $L \sim 0.04 L_{\rm Edd}$),
whereas the coherence of the upper kHz QPO, $Q^{\rm u}_{\rm max}$, is
independent of all the other parameters.

%%%%%%%%%%%%%%%%%%%%%%%%%%%%%%%%%%%%%%%%%%%%%%%%%%%%%%%%%%%%%%%%%%%%%%%%%%%%%%%%%%%%%
%%%%%%%%%%%%%%%%%%%%%%%%%%%%%%%%%%%%%%%%%%%%%%%%%%%%%%%%%%%%%%%%%%%%%%%%%%%%%%%%%%%%%
%%
%% DISCUSSION
%%
%%%%%%%%%%%%%%%%%%%%%%%%%%%%%%%%%%%%%%%%%%%%%%%%%%%%%%%%%%%%%%%%%%%%%%%%%%%%%%%%%%%%%
%%%%%%%%%%%%%%%%%%%%%%%%%%%%%%%%%%%%%%%%%%%%%%%%%%%%%%%%%%%%%%%%%%%%%%%%%%%%%%%%%%%%%

\section{Discussion}
\label{discussion}

%%%%%%%%%%%%%%%%%%%%%%%%%%%%%%%%%%%%%%%%%%%%%%%%%%%%%%%%%%%%%%%%%%%%%%%%%%%%%%%%%%%%%
%%%%%%%%%%%%%%%%%%%%%%%%%%%%%%%%%%%%%%%%%%%%%%%%%%%%%%%%%%%%%%%%%%%%%%%%%%%%%%%%%%%%%
%%
%% FIG. 7
%%
%%%%%%%%%%%%%%%%%%%%%%%%%%%%%%%%%%%%%%%%%%%%%%%%%%%%%%%%%%%%%%%%%%%%%%%%%%%%%%%%%%%%%
%%%%%%%%%%%%%%%%%%%%%%%%%%%%%%%%%%%%%%%%%%%%%%%%%%%%%%%%%%%%%%%%%%%%%%%%%%%%%%%%%%%%%

\begin{figure*}
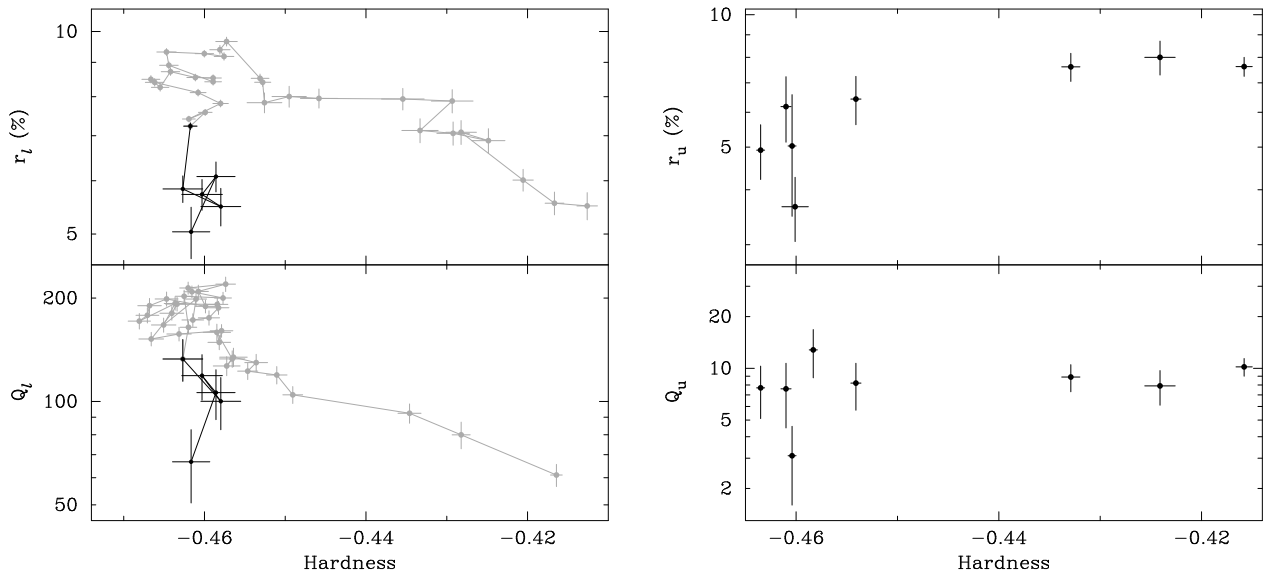

\centerline{\epsfig{file=1608_l.ps,angle=-90,width=8.0cm}~~~~~~
            \epsfig{file=1608_u.ps,angle=-90,width=8.0cm}}
\caption{The rms amplitude (upper panels) and coherence (lower panels)
of the lower kHz QPO (left panels) and the upper kHz QPO (right panels)
for 4U 1608--52 as a function of the hardness of the source. The
hardness is defined as the logarithm of the ratio of the count rate in
the $9.7 - 16.0$ keV band to the count rates in the $6.4 - 9.7$ band
\citep[see][]{mendez-1608}. The lines that connects the points show the
evolution of QPO frequencies, which generally increases from right to
left in these plots.
\label{1608}}
\end{figure*}

%%%%%%%%%%%%%%%%%%%%%%%%%%%%%%%%%%%%%%%%%%%%%%%%%%%%%%%%%%%%%%%%%%%%%%%%%%%%%%%%%%%%%
%%%%%%%%%%%%%%%%%%%%%%%%%%%%%%%%%%%%%%%%%%%%%%%%%%%%%%%%%%%%%%%%%%%%%%%%%%%%%%%%%%%%%
%%
%% SUMMARY OF RESULTS
%%
%%%%%%%%%%%%%%%%%%%%%%%%%%%%%%%%%%%%%%%%%%%%%%%%%%%%%%%%%%%%%%%%%%%%%%%%%%%%%%%%%%%%%
%%%%%%%%%%%%%%%%%%%%%%%%%%%%%%%%%%%%%%%%%%%%%%%%%%%%%%%%%%%%%%%%%%%%%%%%%%%%%%%%%%%%%

I study the maximum amplitude, $r_{\rm max}$, and maximum coherence,
$Q_{\rm max}$, of the kHz QPOs as a function of luminosity and hardness
for a large sample of low-mass X-ray binaries. I show, for the first
time, that the maximum coherence of the lower kHz QPO, $Q^{\ell}_{\rm
max}$, first increases up to $L \sim 0.04 L_{\rm Edd}$ and then
decreases with luminosity, whereas the maximum coherence of the upper
kHz QPO, $Q^{\rm u}_{\rm max}$, is independent of luminosity. I also
find that the maximum rms amplitudes of both the lower and the upper kHz
QPOs, $r^{\ell}_{\rm max}$ and $r^{\rm u}_{\rm max}$, respectively,
decrease monotonically with luminosity and increase monotonically with
the hardness of the source. \citep[The dependence of                    
$r^{\rm u}_{\rm max}$ on luminosity and hardness was first reported
by][]{jonker-0918}.

%%%%%%%%%%%%%%%%%%%%%%%%%%%%%%%%%%%%%%%%%%%%%%%%%%%%%%%%%%%%%%%%%%%%%%%%%%%%%%%%%%%%%
%%%%%%%%%%%%%%%%%%%%%%%%%%%%%%%%%%%%%%%%%%%%%%%%%%%%%%%%%%%%%%%%%%%%%%%%%%%%%%%%%%%%%
%%
%% R-Q correlations
%%
%%%%%%%%%%%%%%%%%%%%%%%%%%%%%%%%%%%%%%%%%%%%%%%%%%%%%%%%%%%%%%%%%%%%%%%%%%%%%%%%%%%%%
%%%%%%%%%%%%%%%%%%%%%%%%%%%%%%%%%%%%%%%%%%%%%%%%%%%%%%%%%%%%%%%%%%%%%%%%%%%%%%%%%%%%%

From the above results it follows that for all sources, $r^{\ell}_{\rm
max}$ and $r^{\rm u}_{\rm max}$ are positively correlated with each
other. Also, for all sources with $L \simmore 0.04 L_{\rm Edd}$, that is
all sources in this paper except 4U 0614+09, the hardest source in the
sample and the one at the lowest luminosity, $Q^{\ell}_{\rm max}$ is
positively correlated both with $r^{\ell}_{\rm max}$ and $r^{\rm u}_{\rm
max}$. $Q^{\rm u}_{\rm max}$ is independent of $Q^{\ell}_{\rm max}$ or
the maximum rms amplitude of the kHz QPOs.

%%%%%%%%%%%%%%%%%%%%%%%%%%%%%%%%%%%%%%%%%%%%%%%%%%%%%%%%%%%%%%%%%%%%%%%%%%%%%%%%%%%%%
%%%%%%%%%%%%%%%%%%%%%%%%%%%%%%%%%%%%%%%%%%%%%%%%%%%%%%%%%%%%%%%%%%%%%%%%%%%%%%%%%%%%%
%%
%% LIFETIME OF THE QPOs
%%
%%%%%%%%%%%%%%%%%%%%%%%%%%%%%%%%%%%%%%%%%%%%%%%%%%%%%%%%%%%%%%%%%%%%%%%%%%%%%%%%%%%%%
%%%%%%%%%%%%%%%%%%%%%%%%%%%%%%%%%%%%%%%%%%%%%%%%%%%%%%%%%%%%%%%%%%%%%%%%%%%%%%%%%%%%%

Since the frequencies at which the $Q^{\ell}_{\rm max}$ values occur are
more or less the same in all sources, and the same is true for $Q^{\rm
u}_{\rm max}$, although the frequencies in this case are higher, the
dependence of $Q^{\ell}_{\rm max}$ and $Q^{\rm u}_{\rm max}$ on
luminosity also reflects the dependence of the maximum lifetime of the
QPOs on luminosity (and spectral hardness). Therefore, for the lower kHz
QPO, the maximum QPO lifetime first increases and then decreases with
luminosity. For the upper kHz QPO, the maximum QPO lifetime is
independent of luminosity.

\subsection{The relation between the ISCO and the drop of QPO coherence
and rms amplitude}  
\label{isco}

%%%%%%%%%%%%%%%%%%%%%%%%%%%%%%%%%%%%%%%%%%%%%%%%%%%%%%%%%%%%%%%%%%%%%%%%%%%%%%%%%%%%%
%%%%%%%%%%%%%%%%%%%%%%%%%%%%%%%%%%%%%%%%%%%%%%%%%%%%%%%%%%%%%%%%%%%%%%%%%%%%%%%%%%%%%
%%
%% q AND r IN INDIVIDUAL SOURCES. DESCRIPTION OF BARRET'S RESULTS
%%
%%%%%%%%%%%%%%%%%%%%%%%%%%%%%%%%%%%%%%%%%%%%%%%%%%%%%%%%%%%%%%%%%%%%%%%%%%%%%%%%%%%%%
%%%%%%%%%%%%%%%%%%%%%%%%%%%%%%%%%%%%%%%%%%%%%%%%%%%%%%%%%%%%%%%%%%%%%%%%%%%%%%%%%%%%%

In individual sources, both $r_{\ell}$ and $Q_{\ell}$ increase with
$\nu_{\ell}$ and then drop rather abruptly at the high end of the
$\nu_{\ell}$ range; $r_{\rm u}$ also increases and then drops at high
$\nu_{\rm u}$ values, and $Q_{\rm u}$ is more or less constant or
increases slightly with $\nu_{\rm u}$ \cite[e.g.,][]{disalvo-1728,
disalvo-1636, mendez-3srcs, vanstraaten-0614-1728, vanstraaten-1608,
barret-1608, barret-1636, barret-nordita, altamirano-1820,
altamirano-1636}. In the case of 4U 1636--53, \cite{barret-1636}
interpret the sudden drop of the coherence and rms amplitude of
$L_{\ell}$, together with the existence of a frequency above which
$L_{\ell}$ is not detected, as evidence of the innermost stable circular
orbit, ISCO, around the neutron star in this system. In general
relativity, different from the Newtonian theory of gravitation, the
effective potential as a function of radial distance to the central
source has a maximum. A particle in a circular orbit at that radius
would be in unstable equilibrium; if perturbed, the particle would fall
onto the central object. No stable orbit around the central object is
possible inside the radius of the ISCO, which in the Schwarzschild case
(non-rotating central object) is $r_{\rm ISCO} = 6 G M/c^2$
\citep*{bardeen}.

%%%%%%%%%%%%%%%%%%%%%%%%%%%%%%%%%%%%%%%%%%%%%%%%%%%%%%%%%%%%%%%%%%%%%%%%%%%%%%%%%%%%%
%%%%%%%%%%%%%%%%%%%%%%%%%%%%%%%%%%%%%%%%%%%%%%%%%%%%%%%%%%%%%%%%%%%%%%%%%%%%%%%%%%%%%
%%
%% COMPARISON BETWEEN INDIVIDUAL SOURCES AND THE SAMPLE
%%
%%%%%%%%%%%%%%%%%%%%%%%%%%%%%%%%%%%%%%%%%%%%%%%%%%%%%%%%%%%%%%%%%%%%%%%%%%%%%%%%%%%%%
%%%%%%%%%%%%%%%%%%%%%%%%%%%%%%%%%%%%%%%%%%%%%%%%%%%%%%%%%%%%%%%%%%%%%%%%%%%%%%%%%%%%%

From the results of individual sources (see references above) and those
of the sample of sources that I present in this paper, it is apparent
that the behaviour of the coherence and rms amplitude of the kHz QPOs as
a function of the {\em QPO frequency} in {\em individual sources} is
similar to the behaviour of the maximum coherence and maximum rms
amplitude of the kHz QPOs as a function of {\em luminosity} in the {\em
sample of sources}. Since in individual sources there is a general
relation between QPO frequencies and source intensity, in the sense that
at higher intensity the QPOs generally appear at higher frequencies (but
remember the parallel-track effect), this raises the question of whether
the same mechanism may be behind both behaviours.

%%%%%%%%%%%%%%%%%%%%%%%%%%%%%%%%%%%%%%%%%%%%%%%%%%%%%%%%%%%%%%%%%%%%%%%%%%%%%%%%%%%%%
%%%%%%%%%%%%%%%%%%%%%%%%%%%%%%%%%%%%%%%%%%%%%%%%%%%%%%%%%%%%%%%%%%%%%%%%%%%%%%%%%%%%%
%%
%% REFERENCE TO FIG. 7
%%
%%%%%%%%%%%%%%%%%%%%%%%%%%%%%%%%%%%%%%%%%%%%%%%%%%%%%%%%%%%%%%%%%%%%%%%%%%%%%%%%%%%%%
%%%%%%%%%%%%%%%%%%%%%%%%%%%%%%%%%%%%%%%%%%%%%%%%%%%%%%%%%%%%%%%%%%%%%%%%%%%%%%%%%%%%%

At first sight, this may appear problematic because it is known that the
relation between QPO frequency and intensity in individual sources is
more complex than the general frequency-luminosity trend described
above, and detailed QPO frequency vs. intensity plots show the so-called
parallel-track phenomenon \citep[see e.g.,][]{mendez-1608}. However, the
link between these two behaviours need not be luminosity, but could be
the high-energy emission (or hardness) in these systems. On one hand, in
the sample of sources the maximum coherence and maximum rms amplitude of
the kHz QPO, except the maximum rms amplitude of $L_{\rm u}$, appear to
correlate fairly well with spectral hardness (see Figure \ref{h1}),
while on the other hand in individual sources the QPO frequencies are
well correlated with the spectral hardness of the source
\citep{mendez-1608}, the index of the power law that fits the
high-energy part of the X-ray spectrum \citep{kaaret-0614-1608}, or $S$,
a parameter that measures the position along the track traced out by the
source in a colour-colour or colour-intensity diagram \citep[called
$S_z$ and $S_a$ in the Z and Atoll sources, respectively; see
e.g.,][]{hertz92, jonker-340+0, mendez-1728}. From this, it follows that
in individual sources there should be a relation between QPO rms
amplitude and coherence on one hand and spectral hardness on the other.
To my knowledge, a plot like that has never been published, but can be
easily constructed using published results. For instance, in the case of
4U 1608--52, this can be done by combining the plots of QPO frequency
vs. hardness of Figure 3 in \cite{mendez-1608} and the plots of QPO
coherence and QPO rms amplitude vs. QPO frequency of Figure 2 in
\cite{barret-nordita} and Figure 3 in \cite{mendez-3srcs}, respectively.
In Figure \ref{1608} I show the rms amplitude (upper panel) and
coherence (lower panel) of the lower kHz QPO (left panel) and the upper
kHz QPO (right panel) in 4U 1608--52 as a function of hardness. In this
case the hardness is defined \citep[see][]{mendez-1608} as the logarithm
of the ratio of the PCA count rate in the $9.7 - 16.0$ keV band to the
PCA count rates in the $6.4 - 9.7$ band. (Notice that the hardness in
this plot is defined over a different energy band than in the case shown
in Figure \ref{h1}.) The lines connecting the data points indicate the
evolution of QPO frequency, which generally increases from right to
left. This Figure shows the drop of the rms amplitude and coherence of
the lower kHz QPO and of the rms amplitude of the upper kHz QPO at the
very high end of the QPO frequency range \citep[cf][]{disalvo-1728,
disalvo-1636, mendez-3srcs, barret-1608, barret-1636, barret-nordita}

%%%%%%%%%%%%%%%%%%%%%%%%%%%%%%%%%%%%%%%%%%%%%%%%%%%%%%%%%%%%%%%%%%%%%%%%%%%%%%%%%%%%%
%%%%%%%%%%%%%%%%%%%%%%%%%%%%%%%%%%%%%%%%%%%%%%%%%%%%%%%%%%%%%%%%%%%%%%%%%%%%%%%%%%%%%
%%
%% COMPARISON BETWEEN INDIVIDUAL SOURCES (USING THE CASE OF 1608-52) AND
%% THE SAMPLE
%%
%%%%%%%%%%%%%%%%%%%%%%%%%%%%%%%%%%%%%%%%%%%%%%%%%%%%%%%%%%%%%%%%%%%%%%%%%%%%%%%%%%%%%
%%%%%%%%%%%%%%%%%%%%%%%%%%%%%%%%%%%%%%%%%%%%%%%%%%%%%%%%%%%%%%%%%%%%%%%%%%%%%%%%%%%%%

The trend seen in Figure \ref{1608} for the coherence of the lower kHz
QPO in 4U 1608--52 is rather similar to the behaviour of the maximum
coherence of the lower kHz QPO for the sample of sources in Figure
\ref{h1} (left panel; notice again that the hardness is defined over a
different energy band in each Figure): When 4U 1608--52 is hard, to the
right of Figure \ref{1608}, the coherence is low, while for the hardest
source in the sample, to the right of the left panel of Figure \ref{h1},
the maximum coherence is also low. When 4U 1608--52 becomes softer, to
the left of Figure \ref{1608}, the coherence first increases, and then
drops abruptly (see black points in Figure \ref{1608}, lower panel),
while in the sample of sources the maximum coherence behaves in a
similar way. The behaviour of the rms amplitude in 4U 1608--52 and the
maximum rms amplitude in the sample of sources is more or less similar,
except that in the case of 4U 1608--52 when the source is hard the rms
amplitude is low, increases as the source gets softer and then drops
abruptly (see black points in Figure \ref{1608}, upper panel), whereas
in the sample of sources moving from hard to soft sources the maximum
rms amplitude remains more or less constant and at the end drops. To
summarize, the relevant conclusion from this comparison is this:  Both
the rms amplitude and coherence of the lower kHz QPO in 4U 1608--52 and
the maximum rms amplitude and maximum coherence of the lower kHz QPO in
the sample of sources drop abruptly when the hardness decreases.

%%%%%%%%%%%%%%%%%%%%%%%%%%%%%%%%%%%%%%%%%%%%%%%%%%%%%%%%%%%%%%%%%%%%%%%%%%%%%%%%%%%%%
%%%%%%%%%%%%%%%%%%%%%%%%%%%%%%%%%%%%%%%%%%%%%%%%%%%%%%%%%%%%%%%%%%%%%%%%%%%%%%%%%%%%%
%%
%% SIMILAR TRENDS IN INDIVIDUAL SOURCES AND THE SAMPLE SUGGEST SAME
%% UNDERLYING MECHANISM
%%
%%%%%%%%%%%%%%%%%%%%%%%%%%%%%%%%%%%%%%%%%%%%%%%%%%%%%%%%%%%%%%%%%%%%%%%%%%%%%%%%%%%%%
%%%%%%%%%%%%%%%%%%%%%%%%%%%%%%%%%%%%%%%%%%%%%%%%%%%%%%%%%%%%%%%%%%%%%%%%%%%%%%%%%%%%%

The comparison between individual sources and the sample of sources
suggests that the same mechanism is responsible for the drop of
coherence and rms amplitude of the lower kHz QPO with QPO frequency in
individual sources as well as for the drop of maximum QPO coherence and
maximum QPO rms amplitude with luminosity in the sample of sources. The
comparison between 4U 1608--52 and the sample of sources in the previous
paragraph suggests that most likely the mechanism is related to the
high-energy emission in these systems. This does not necessarily mean
that the fractional emission at high energies (represented by the
hardness or X-ray colors) is the root mechanism that drives all QPO
parameters (QPO frequency, coherence, and rms amplitude). For instance,
one possibility (there could be many others) is that the (instantaneous)
mass accretion rate sets the size of the inner radius of the disc
\citep{vanderklis-2001}, which in turn determines the QPO frequency, as
well as the relative contribution of the high-energy part of the
spectrum to the total luminosity. If the efficiency of the modulation
mechanism and the lifetime of the oscillations that produce the QPO
depended upon the emission from the high-energy part of the source
spectrum (see \S~\ref{modulation} for a discussion of possible ways in
which this could happen), observationally it would appear as if the
coherence and rms amplitude of the QPO depended upon the QPO frequency,
and hence upon the radius in the disc at which the QPO is produced. The
sudden drop of the coherence and rms amplitude of the QPO at some QPO
frequency would then be associated to a dynamical peculiarity in the
accretion disc, for instance the ISCO. Observing the same source
repeatedly would not allow to distinguish the above scenario from one in
which QPO coherence and rms amplitude were actually set by QPO frequency
or the dynamics in the accretion disc.

To distinguish between these two options, one would need to observe a
sample of sources of kHz QPOs for which the mass-accretion rate, and
hence the relative contribution of the high-energy part of the spectrum
to the total emission, was different from source to source. In that
case, QPO coherence and rms amplitude would drop for sources accreting
mass at higher rates, even if the frequency of the QPO was more or less
the same from source to source. Since, as I have shown in this paper,
this is the general behaviour observed in sources of kHz QPOs, it is
reasonable to infer that a mechanism similar to the one I described in
the previous paragraph is effective in setting the coherence and rms
amplitude of the kHz QPOs. If this is correct, this also implies that
the drop of QPO coherence and rms amplitude as a function of QPO
frequency in individual sources cannot be due to effects of the ISCO.

%%%%%%%%%%%%%%%%%%%%%%%%%%%%%%%%%%%%%%%%%%%%%%%%%%%%%%%%%%%%%%%%%%%%%%%%%%%%%%%%%%%%%
%%%%%%%%%%%%%%%%%%%%%%%%%%%%%%%%%%%%%%%%%%%%%%%%%%%%%%%%%%%%%%%%%%%%%%%%%%%%%%%%%%%%%
%%
%% RMS OF OTHER QPOS ALSO DROP WITH NU_U
%%
%%%%%%%%%%%%%%%%%%%%%%%%%%%%%%%%%%%%%%%%%%%%%%%%%%%%%%%%%%%%%%%%%%%%%%%%%%%%%%%%%%%%%
%%%%%%%%%%%%%%%%%%%%%%%%%%%%%%%%%%%%%%%%%%%%%%%%%%%%%%%%%%%%%%%%%%%%%%%%%%%%%%%%%%%%%

Note also that in individual sources not just the rms amplitude of the
kHz QPOs, but also the rms amplitude of other lower-frequency QPOs
decrease with increasing QPO frequency. For instance, in four Atoll
sources, 4U 1728--34, 4U 1608--52, 4U 0614+09, and 4U 1636--53, the rms
amplitudes of the ``bump'', a QPO at $\sim 0.1-30$ Hz, the ``hump'', a
QPO at $\sim 1-40$, and the hectohertz QPO at $\sim 100-300$ Hz, all
drop as the frequencies of the kHz QPOs increase, in a similar fashion
as the amplitude of the upper and lower kHz QPOs \citep[e.g.,][and
references therein; see also there a description of these other
QPOs]{altamirano-1636}. This also argues against the interpretation of
the ISCO as the cause of the drop of the rms of the kHz QPOs, and
indicates that the amplitudes of {\em all} variability components are
set by the same mechanism which, as I suggested, could be the same one
that governs the high-energy spectral component.

%%%%%%%%%%%%%%%%%%%%%%%%%%%%%%%%%%%%%%%%%%%%%%%%%%%%%%%%%%%%%%%%%%%%%%%%%%%%%%%%%%%%%
%%%%%%%%%%%%%%%%%%%%%%%%%%%%%%%%%%%%%%%%%%%%%%%%%%%%%%%%%%%%%%%%%%%%%%%%%%%%%%%%%%%%%
%%
%% POTENTIAL TEST OF THE IDEA THAT RMS AND Q DEPEND ON HIGH-ENERGY
%% EMISSION (OR HARDNESS)
%%
%%%%%%%%%%%%%%%%%%%%%%%%%%%%%%%%%%%%%%%%%%%%%%%%%%%%%%%%%%%%%%%%%%%%%%%%%%%%%%%%%%%%%
%%%%%%%%%%%%%%%%%%%%%%%%%%%%%%%%%%%%%%%%%%%%%%%%%%%%%%%%%%%%%%%%%%%%%%%%%%%%%%%%%%%%%

The idea that the rms amplitude and coherence of the QPOs depend on the
high-energy emission, and not on QPO frequency, can in principle be
tested; for instance, if one of the kHz QPO sources showed a sudden
change in the high-energy part of the spectrum, but the change of QPO
frequency was less sudden. This could for example occur if the component
that sets the high-energy emission in the spectrum of these sources and
the one that sets the QPO frequency had different responses to changes
of the mass accretion rate. This scenario is similar to the one proposed
by \cite{vanderklis-2001} to explain the parallel tracks in the
frequency vs. intensity plots of these sources. If this was the case,
one would observe a kHz QPO that had a coherence and rms amplitude that
would not match those expected on the basis of the $Q-\nu$ and $r-\nu$
relations in, e.g., \cite{barret-1608, barret-1636, barret-nordita}.  I
would like to point out that Figure 1 in \cite{barret-4srcs} appears to
show an effect like this: Individual measurements of the coherence of
the lower kHz QPO at the same QPO frequency differ significantly from
one another. E.g., in 4U 1636--53 at $\nu_\ell \approx 820$ Hz, $Q_\ell$
ranges from $Q_\ell = 80 \pm 20$ to $Q_\ell = 180 \pm 10$. This
difference, however, may still be due to uncorrected drifts of the QPO
frequency during the intervals in which \cite{barret-4srcs} measured the
coherence of the QPO, although according to their description they had
taken this into account in producing their Figure.

\subsection{The modulation mechanism and the lifetime of the kHz QPOs}
\label{modulation}

%%%%%%%%%%%%%%%%%%%%%%%%%%%%%%%%%%%%%%%%%%%%%%%%%%%%%%%%%%%%%%%%%%%%%%%%%%%%%%%%%%%%%
%%%%%%%%%%%%%%%%%%%%%%%%%%%%%%%%%%%%%%%%%%%%%%%%%%%%%%%%%%%%%%%%%%%%%%%%%%%%%%%%%%%%%
%%
%% MECHANISM: OSCILLATION IN THE DISK, MODULATION ELSEWHERE
%%
%%%%%%%%%%%%%%%%%%%%%%%%%%%%%%%%%%%%%%%%%%%%%%%%%%%%%%%%%%%%%%%%%%%%%%%%%%%%%%%%%%%%%
%%%%%%%%%%%%%%%%%%%%%%%%%%%%%%%%%%%%%%%%%%%%%%%%%%%%%%%%%%%%%%%%%%%%%%%%%%%%%%%%%%%%%

How can the mechanism that produces the high-energy emission in these
systems change the coherence and rms amplitude of the kHz QPOs? There
is general agreement that the oscillation mechanism that produces the
quasi-periodic variability must be in the disk (see the references to
theoretical models in \S \ref{intro}); it is very easy to come up with
characteristic dynamical frequencies in the disk that match the observed
frequencies of the kHz QPOs, whereas it appears more difficult to have a
``clock'' somewhere else in the accretion flow. Nevertheless, from the
early observations it became clear that emission from the disk alone
cannot explain the rms amplitudes of the kHz QPOs, since in some cases
the modulated luminosity in the QPOs is $\sim 15\%$ of the total
emission, whereas at the same time the emission of the disk is $\sim
10\%$ or less. Furthermore, the steep increase of the rms amplitude with
energy imply a large modulation of the emitted flux, up to $\sim 20\%$
at $\sim 25-30$ keV \citep[see e.g.,][]{berger-1608}, at energies where
the contribution of the disk is negligible. Hence, while the oscillation
mechanism probably takes place in the disk, most likely the modulation
mechanism is associated to the high-energy spectral component in these
sources.

%%%%%%%%%%%%%%%%%%%%%%%%%%%%%%%%%%%%%%%%%%%%%%%%%%%%%%%%%%%%%%%%%%%%%%%%%%%%%%%%%%%%%
%%%%%%%%%%%%%%%%%%%%%%%%%%%%%%%%%%%%%%%%%%%%%%%%%%%%%%%%%%%%%%%%%%%%%%%%%%%%%%%%%%%%%
%%
%% HIGH-ENERGY SPECTRUM OF LMXBS; CHANGES OF PARAMETERS OF THE
%% COMPTONIZING COMPONENT (KT_E AND TAU) WITH SOURCE STATE AND ACROSS 
%% SOURCES
%%
%%%%%%%%%%%%%%%%%%%%%%%%%%%%%%%%%%%%%%%%%%%%%%%%%%%%%%%%%%%%%%%%%%%%%%%%%%%%%%%%%%%%%
%%%%%%%%%%%%%%%%%%%%%%%%%%%%%%%%%%%%%%%%%%%%%%%%%%%%%%%%%%%%%%%%%%%%%%%%%%%%%%%%%%%%%

The high-energy part of the X-ray spectrum of low-mass X-ray binaries
can be described in terms of thermal Comptonization
\citep*{sunyaev-titarchuk, white-stella-parmar}. In low-luminosity Atoll
sources \citep{hk89}, the hard spectral changes are explained as due to
changes of the properties of this component as the average
mass-accretion rate, $\dot M$, changes. For instance, for 4U 1608--52,
\cite{gierlinski-1608} find that at low inferred $\dot M$, the
Comptonizing component has a relatively high temperature, $T_e \simmore
20$ keV, and is optically thin, $\tau_e \sim 1$. The high-energy part of
the spectrum resembles a rather flat power law (power-law index $\sim
1.5-2$), with a high-energy cut-off above $\sim 100$ keV
\citep{zhang-1608}. At high inferred $\dot M$, \cite{gierlinski-1608}
find that the Comptonizing component becomes cooler, with $T_e \simless
5$ keV and the optical depth increases, $\tau_e >> 1$. The high-energy
part of the spectrum resembles a power law that is steeper (power-law
index $\sim 2-2.5$) with a high-energy cut-off that is at lower energies
than in the low $\dot M$ case. In the case of high-luminosity Z-sources
\citep{hk89}, the Comptonizing component is relatively cool and
optically thick as in the case of the low-luminosity Atoll sources at
high $\dot M$ \citep{christian-swank}.

%%%%%%%%%%%%%%%%%%%%%%%%%%%%%%%%%%%%%%%%%%%%%%%%%%%%%%%%%%%%%%%%%%%%%%%%%%%%%%%%%%%%%
%%%%%%%%%%%%%%%%%%%%%%%%%%%%%%%%%%%%%%%%%%%%%%%%%%%%%%%%%%%%%%%%%%%%%%%%%%%%%%%%%%%%%
%%
%% MODEL OF VARIABILITY BY LEE AND MILLER
%%
%%%%%%%%%%%%%%%%%%%%%%%%%%%%%%%%%%%%%%%%%%%%%%%%%%%%%%%%%%%%%%%%%%%%%%%%%%%%%%%%%%%%%
%%%%%%%%%%%%%%%%%%%%%%%%%%%%%%%%%%%%%%%%%%%%%%%%%%%%%%%%%%%%%%%%%%%%%%%%%%%%%%%%%%%%%

Using a simple time-dependent Comptonization model, \cite{lee-miller}
calculated the spectrum of variability that would be produced by
oscillations in the (i) injection rate of seed photons, (ii) density,
and (iii) temperature of the Comptonizing medium \citep[see
also][]{stollman}. They find that to reproduce the rms spectrum of the
lower kHz QPO in 4U 1608--52, the variability must be mostly driven by
an oscillation of the density of the Comptonizing medium. They also find
that in the case of variations of the density of the Comptonizing
medium, the largest rms variability in the $1-10$ keV range occurs as
the optical depth of the medium is smallest (see their Figure 4, panel
b).

%%%%%%%%%%%%%%%%%%%%%%%%%%%%%%%%%%%%%%%%%%%%%%%%%%%%%%%%%%%%%%%%%%%%%%%%%%%%%%%%%%%%%
%%%%%%%%%%%%%%%%%%%%%%%%%%%%%%%%%%%%%%%%%%%%%%%%%%%%%%%%%%%%%%%%%%%%%%%%%%%%%%%%%%%%%
%%
%% QUALITATIVE EXPLANATION OF THE RESULTS
%%
%%%%%%%%%%%%%%%%%%%%%%%%%%%%%%%%%%%%%%%%%%%%%%%%%%%%%%%%%%%%%%%%%%%%%%%%%%%%%%%%%%%%%
%%%%%%%%%%%%%%%%%%%%%%%%%%%%%%%%%%%%%%%%%%%%%%%%%%%%%%%%%%%%%%%%%%%%%%%%%%%%%%%%%%%%%

If the high-energy emission is due to Comptonization, the results of
\cite{lee-miller}, together with the global correlation between
luminosity and spectral hardness \citep{vanparadijs-vanderklis}, provide
a possible explanation, at least qualitative, of the dependence of the
rms amplitude of the kHz QPOs on luminosity and hardness. At low
luminosity, corresponding to low mass accretion rate and hard spectra,
the Comptonizing plasma is optically thin, and hence the amplitude of
the variability is high. When mass accretion rate increases, the
luminosity increases and the source becomes softer, and since the
optical depth of the Comptonizing plasma increases, the amplitude of the
variability decreases. This would explain both the drop in rms amplitude
of the kHz QPOs at high QPO frequency in individual sources, and that of
the maximum rms amplitude in the sample of sources at high luminosity
(low hardness). In individual sources, the rms amplitude of the kHz QPOs
also decreases at low frequencies, which is more difficult to explain in
this simplified scenario. Also, in this scenario there is no
straightforward explanation of the drop of the coherence of the lower
kHz QPO at high frequencies in individual sources, and at high
luminosities in the sample of sources.

%%%%%%%%%%%%%%%%%%%%%%%%%%%%%%%%%%%%%%%%%%%%%%%%%%%%%%%%%%%%%%%%%%%%%%%%%%%%%%%%%%%%%
%%%%%%%%%%%%%%%%%%%%%%%%%%%%%%%%%%%%%%%%%%%%%%%%%%%%%%%%%%%%%%%%%%%%%%%%%%%%%%%%%%%%%
%%
%% EXPLANATION VIA DE BOUNDARY LAYER
%%
%%%%%%%%%%%%%%%%%%%%%%%%%%%%%%%%%%%%%%%%%%%%%%%%%%%%%%%%%%%%%%%%%%%%%%%%%%%%%%%%%%%%%
%%%%%%%%%%%%%%%%%%%%%%%%%%%%%%%%%%%%%%%%%%%%%%%%%%%%%%%%%%%%%%%%%%%%%%%%%%%%%%%%%%%%%

\cite*{gilfanov-aa} have analyzed the rms energy spectra as a function
of frequency of two low-mass X-ray binaries, the Z source GX 340+0 and
the Atoll source 4U 1608--52. They find that the variability, including
that in the frequency range of the kHz QPOs, is primarily due to
variations of the luminosity of the boundary layer, with the emission of
the accretion disk being much less variable. They also find \citep[see
also][]{gilfanov-nordita} that in GX 340+0, the contribution of the
boundary layer to the observed emission decreases as mass accretion rate
increases, as the source moves along the Z-shaped track in the
color-color diagram from the Horizontal Branch to the Normal Branch
\citep[see][for an explanation of the branches in the color-color
diagram of Z sources]{hk89}. \cite{gilfanov-aa} speculate that this
decrease could be due either to obscuration of the boundary layer by a
thickened accretion disk, a quantitative change in the structure of the
boundary layer, or a complete disappearance of the boundary layer, when
$\dot M \sim \dot M_{\rm Edd}$. In this scenario, the decrease of the
rms variability of the kHz QPOs at high luminosities (high mass
accretion rate) would be related to the lower relative contribution of
the boundary layer, which produces the bulk of the QPO variability, to
the total emission. The obscuration mechanism cannot explain, however,
the decrease of the coherence of the QPO at high luminosity. One way to
explain the loss of coherence would be if the modulation mechanism in
the boundary layer would damp out the oscillations. For instance, the
flow of mass onto the boundary layer could be modulated more or less
periodically by the disc, and emission would proceed as a series of
shots when mass reaches the boundary layer. Changes in the configuration
of the boundary layer, as those proposed by \cite{gilfanov-aa}, could
increase the damping and reduce the lifetime of the QPO. This is of
course just speculation; it remains to be seen whether it is possible to
explain the decrease of QPO coherence by significant changes in the
properties of the boundary layer.

%%%%%%%%%%%%%%%%%%%%%%%%%%%%%%%%%%%%%%%%%%%%%%%%%%%%%%%%%%%%%%%%%%%%%%%%%%%%%%%%%%%%%
%%%%%%%%%%%%%%%%%%%%%%%%%%%%%%%%%%%%%%%%%%%%%%%%%%%%%%%%%%%%%%%%%%%%%%%%%%%%%%%%%%%%%
%%
%% PRESENCE/ABSENCE OF THE POWER LAW EMISSION AS A FUNCTION OF L_X
%% (ACCORDING TO BARRET & VERDENNE)
%%
%%%%%%%%%%%%%%%%%%%%%%%%%%%%%%%%%%%%%%%%%%%%%%%%%%%%%%%%%%%%%%%%%%%%%%%%%%%%%%%%%%%%%
%%%%%%%%%%%%%%%%%%%%%%%%%%%%%%%%%%%%%%%%%%%%%%%%%%%%%%%%%%%%%%%%%%%%%%%%%%%%%%%%%%%%%

\cite{barret-verdenne} suggested that there is a critical luminosity for
low-mass X-ray binaries with a neutron-star primary, $L \sim 10^{36} -
10^{37}$ erg s$^{-1}$; sources below that level show a hard power-law
component, whereas sources above that that level do not. Figure
\ref{low} shows a sort of dichotomy between the maximum rms amplitude
and maximum coherence of the kHz QPOs in the Z sources, for which $L
\simmore 0.8 - 1.0 L_{\rm Edd}$ and the Atoll sources, for which $L
\simless 0.1 - 0.2 L_{\rm Edd}$. The separation, however, occurs at a
luminosity that is $\sim 5 -10$ times larger than suggested by
\cite{barret-verdenne}. Nevertheless, at least for the lower kHz QPO,
the maximum coherence and maximum rms amplitude change significantly and
rather smoothly within the Atoll class; this may indicate that, if the
QPO properties depend upon the emission of the power-law component, the
relative importance of this component may change in a less abrupt manner
than suggested by \cite{barret-verdenne}.

%%%%%%%%%%%%%%%%%%%%%%%%%%%%%%%%%%%%%%%%%%%%%%%%%%%%%%%%%%%%%%%%%%%%%%%%%%%%%%%%%%%%%
%%%%%%%%%%%%%%%%%%%%%%%%%%%%%%%%%%%%%%%%%%%%%%%%%%%%%%%%%%%%%%%%%%%%%%%%%%%%%%%%%%%%%
%%
%% CORRELATION BETWEEN RMS AND Q SUGGEST LINKED MECHANISM IN SOME CASES
%%
%%%%%%%%%%%%%%%%%%%%%%%%%%%%%%%%%%%%%%%%%%%%%%%%%%%%%%%%%%%%%%%%%%%%%%%%%%%%%%%%%%%%%
%%%%%%%%%%%%%%%%%%%%%%%%%%%%%%%%%%%%%%%%%%%%%%%%%%%%%%%%%%%%%%%%%%%%%%%%%%%%%%%%%%%%%

The $r^{\ell}_{\rm max} - r^{\rm u}_{\rm max}$ correlation
(Fig.~\ref{cross-plots}) suggests that the same mechanism sets the
amplitude of both QPOs. But in each source $r^{\ell}_{\rm max}$ and
$r^{\rm u}_{\rm max}$ occur at different luminosities (see Tables
\ref{table_l} and \ref{table_u}), and hence at different times,
therefore the mechanism cannot be acting simultaneously on both kHz
QPOs. \citep[In fact, in each source $r^{\ell}_{\rm max}$ occurs at a
higher frequency of the upper kHz QPO, $\nu_{\rm u}$ than $r^{\rm
u}_{\rm max}$; see, e.g.,][]{vanstraaten-0614-1728, vanstraaten-1608}.
If one ignores the case of 4U 0614+09, the source with the lowest
luminosity in the sample, also $Q^{\ell}_{\rm max}$ is correlated with
$r^{\ell}_{\rm max}$ as well as with $r^{\rm u}_{\rm max}$. From this it
is tempting to infer that the mechanism that sets the rms amplitude of
the kHz QPOs sets the coherence of the lower kHz QPO as well, at least
for $L \simmore 0.04 L_{\rm Edd}$. At low luminosities either this
mechanism is not effective in setting $Q^{\ell}_{\rm max}$, or another
mechanism comes into play that counteracts the effect. The lack of a
correlation between $Q^{\rm u}_{\rm max}$ and $Q^{\ell}_{\rm max}$,
$r^{\ell}_{\rm max}$, or $r^{\rm u}_{\rm max}$ indicates that the
mechanism that sets the latter three quantities does not have a
significant effect on the coherence of the upper kHz QPO.

%%%%%%%%%%%%%%%%%%%%%%%%%%%%%%%%%%%%%%%%%%%%%%%%%%%%%%%%%%%%%%%%%%%%%%%%%%%%%%%%%%%%%
%%%%%%%%%%%%%%%%%%%%%%%%%%%%%%%%%%%%%%%%%%%%%%%%%%%%%%%%%%%%%%%%%%%%%%%%%%%%%%%%%%%%%
%%
%% Q_MAX FOR L_\ELL IN 0614 IS TOO LOW
%%
%%%%%%%%%%%%%%%%%%%%%%%%%%%%%%%%%%%%%%%%%%%%%%%%%%%%%%%%%%%%%%%%%%%%%%%%%%%%%%%%%%%%%
%%%%%%%%%%%%%%%%%%%%%%%%%%%%%%%%%%%%%%%%%%%%%%%%%%%%%%%%%%%%%%%%%%%%%%%%%%%%%%%%%%%%%

As I showed in \S\ref{results}, for 4U 0614+09 either $Q^{\ell}_{\rm
max}$ is too low for $r^{\ell}_{\rm max}$, or $r^{\ell}_{\rm max}$ is
too high for $Q^{\ell}_{\rm max}$. But since the maximum coherence for
$L_{\ell}$ in 4U 0614+09 lies away from the $r^{\ell}_{\rm max} -
Q^{\ell}_{\rm max}$ as well as from the $r^{\rm u}_{\rm max} -
Q^{\ell}_{\rm max}$ correlations, it appears more likely that in this
case $Q^{\ell}_{\rm max}$ is too low both for the $r^{\ell}_{\rm max}$
and $r^{\rm u}_{\rm max}$, and not the other way around. This hypothesis
is reinforced by the fact that a larger $Q^{\ell}_{\rm max}$ would bring
4U 0614+09 closer to the other Atoll sources in Figures \ref{rmsq} and
\ref{cross-plots}, whereas lower $r^{\ell}_{\rm max}$ and $r^{\rm
u}_{\rm max}$ values would put 4U 0614+09 in the gap of the correlations
in those two Figures. If this is the case, it is unclear what mechanism
could be responsible for the drop of coherence without a simultaneous
drop of rms amplitude.

%%%%%%%%%%%%%%%%%%%%%%%%%%%%%%%%%%%%%%%%%%%%%%%%%%%%%%%%%%%%%%%%%%%%%%%%%%%%%%%%%%%%%
%%%%%%%%%%%%%%%%%%%%%%%%%%%%%%%%%%%%%%%%%%%%%%%%%%%%%%%%%%%%%%%%%%%%%%%%%%%%%%%%%%%%%
%%
%% COMPARISON OF Q_U MAX IN THE POP. AND Q_U IN INDIVIDUAL SOURCES
%%
%%%%%%%%%%%%%%%%%%%%%%%%%%%%%%%%%%%%%%%%%%%%%%%%%%%%%%%%%%%%%%%%%%%%%%%%%%%%%%%%%%%%%
%%%%%%%%%%%%%%%%%%%%%%%%%%%%%%%%%%%%%%%%%%%%%%%%%%%%%%%%%%%%%%%%%%%%%%%%%%%%%%%%%%%%%

Comparison of the upper right panels of Figures \ref{h1} and \ref{1608}
suggests that the behaviors of $Q_{\rm u}$ in individual sources (in
this case 4U 1608--52) and $Q^{\rm u}_{\rm max}$ in the sample of
sources as a function of hardness are also similar, in that both the
coherence of the upper kHz QPO in individual sources, and the maximum
coherence of the upper kHz QPO in the sample of sources are more or less
independent of the hardness of the source. However, the lack of
correlation between $Q_{\rm u}$ and QPO frequency in 4U 1608--52 appears
to be exceptional, and in other sources, both Atoll \citep[see,
e.g.,][]{vanstraaten-0614-1728, altamirano-1636} as well as Z-sources
\citep[see, e.g.,][]{vanderklis-scox-1, homan-17+2, jonker-340+0,
jonker-5-1}, $Q_{\rm u}$ increases with QPO frequency, and hence
decreases with hardness. From the data in Figure \ref{h1} it is not
possible to discard a similar slow decrease of $Q^{\rm u}_{\rm max}$ vs.
hardness.

%%%%%%%%%%%%%%%%%%%%%%%%%%%%%%%%%%%%%%%%%%%%%%%%%%%%%%%%%%%%%%%%%%%%%%%%%%%%%%%%%%%%%
%%%%%%%%%%%%%%%%%%%%%%%%%%%%%%%%%%%%%%%%%%%%%%%%%%%%%%%%%%%%%%%%%%%%%%%%%%%%%%%%%%%%%
%%
%% THE CASE OF BHC
%%
%%%%%%%%%%%%%%%%%%%%%%%%%%%%%%%%%%%%%%%%%%%%%%%%%%%%%%%%%%%%%%%%%%%%%%%%%%%%%%%%%%%%%
%%%%%%%%%%%%%%%%%%%%%%%%%%%%%%%%%%%%%%%%%%%%%%%%%%%%%%%%%%%%%%%%%%%%%%%%%%%%%%%%%%%%%

In black-hole systems, high-frequency QPOs have been observed in the
range of tens up to $\sim 450$ Hz with typical rms amplitude and
coherence values $r \sim 1-3$\% and $Q \sim 2-10$, respectively
\citep*{morgan-1915, remillard-1655, remillard-1550, remillard-1915,
cui-1859, belloni-1915a, strohmayer-1655, strohmayer-1915, homan-1550,
homan-1650, homan-1743, miller-1550, klein-wolt-1630, belloni-1915b}.
The luminosities of these systems when high-frequency QPOs are observed
are $L \sim 1-3 \times 10^{38}$ erg s$^{-1}$ \citep[e.g., in XTE
J1550--564;][]{homan-1550, sobczak-1550} or, for the typical mass of the
black holes, $M_{\rm BH} \sim 10 \msun$, $L \sim 0.1 L_{\rm Edd}$.
Compared to the neutron-star systems in Figure \ref{low}, in black-hole
systems the rms amplitude and coherence of the high-frequency QPOs are
too low for $L/L_{\rm Edd}$ of the source, but they would be consistent
with the values for the Z sources at the same luminosity, i.e., the
luminosities not being normalized to the source own Eddington luminosity
(of course, to plot the black-hole sources in Figure \ref{low} one would
have to divide their luminosities by $2.5 \times 10^{38}
\mathrm{erg}~\mathrm{s}^{-1}$, the Eddington luminosity for a $1.9
\msun$ neutron star, as was done for the other sources in that plot). It
remains to be seen whether in black-hole systems the modulation
mechanism of the high-frequency QPOs would be similar to that of the kHz
QPOs in neutron-star systems.

%%%%%%%%%%%%%%%%%%%%%%%%%%%%%%%%%%%%%%%%%%%%%%%%%%%%%%%%%%%%%%%%%%%%%%%%%%%%%%%%%%%%%
%%%%%%%%%%%%%%%%%%%%%%%%%%%%%%%%%%%%%%%%%%%%%%%%%%%%%%%%%%%%%%%%%%%%%%%%%%%%%%%%%%%%%
%%
%% CONCLUSIONS
%%
%%%%%%%%%%%%%%%%%%%%%%%%%%%%%%%%%%%%%%%%%%%%%%%%%%%%%%%%%%%%%%%%%%%%%%%%%%%%%%%%%%%%%
%%%%%%%%%%%%%%%%%%%%%%%%%%%%%%%%%%%%%%%%%%%%%%%%%%%%%%%%%%%%%%%%%%%%%%%%%%%%%%%%%%%%%

\section{Conclusion}
\label{conclusion}

I study the maximum amplitude, $r_{\rm max}$, and maximum coherence,
$Q_{\rm max}$, of the kHz QPOs as a function of luminosity and hardness
for a dozen low-mass X-ray binaries. I find that:

\begin{enumerate}

\item The maximum coherence of the lower kHz QPO, $Q^{\ell}_{\rm max}$,
first increases up to $L \sim 0.04 L_{\rm Edd}$ and then decreases with
luminosity. 

\item The maximum coherence of the upper kHz QPO, $Q^{\rm u}_{\rm max}$,
is independent of luminosity. 

\item The maximum rms amplitudes of both the lower and the upper kHz
QPOs, $r^{\ell}_{\rm max}$ and $r^{\rm u}_{\rm max}$, respectively,
decrease monotonically with luminosity

\item Both $r^{\rm u}_{\rm max}$ and $r^{\ell}_{\rm max}$ increase with
the source hardness, $Q^{\ell}_{\rm max}$ first increases with hardness
and then drops for the hardest source in the sample, and $Q^{\rm u}_{\rm
max}$ is independent of hardness.

\item The relation between $Q_{\rm max}$ and $r_{\rm max}$ with
luminosity in the sample of sources is similar to the relation between
$Q$ and $r$ with QPO frequency in individual sources. The similarity
extends also to hardness in the sample of sources and in at least one
individual source, 4U 1608--52.

\item The above argues against the interpretation that the drop of QPO
coherence and QPO rms amplitude at high QPO frequency in individual
sources is due to effects related to the innermost stable orbit around
the neutron star in these systems.

\item The drop of coherence and rms amplitude of the kHz QPOs, both in
individual sources and in the sample could be produce by changes in the
properties of the region in the accretion flow where the flux that
produces the QPO is modulated.

\end{enumerate}

%%%%%%%%%%%%%%%%%%%%%%%%%%%%%%%%%%%%%%%%%%%%%%%%%%%%%%%%%%%%%%%%%%%%%%%%%%%%%%%%%%%%%
%%%%%%%%%%%%%%%%%%%%%%%%%%%%%%%%%%%%%%%%%%%%%%%%%%%%%%%%%%%%%%%%%%%%%%%%%%%%%%%%%%%%%
%%
%% ACKNOWLEDGMENTS
%%
%%%%%%%%%%%%%%%%%%%%%%%%%%%%%%%%%%%%%%%%%%%%%%%%%%%%%%%%%%%%%%%%%%%%%%%%%%%%%%%%%%%%%
%%%%%%%%%%%%%%%%%%%%%%%%%%%%%%%%%%%%%%%%%%%%%%%%%%%%%%%%%%%%%%%%%%%%%%%%%%%%%%%%%%%%%

\section*{Acknowledgments}

I thank Didier Barret and Cole Miller for valuable discussions on the
ideas that I present in this paper. I thank Diego Altamirano for
comments on an earlier version of this manuscript. I also thank Michiel
van der Klis for insightful comments that helped improve the paper. This
research has made use of data obtained through the High Energy
Astrophysics Science Archive Research Center Online Service, provided by
the NASA/Goddard Space Flight Center. The Netherlands Institute for
Space Research (SRON) is supported financially by NWO, the Netherlands
Organisation for Scientific Research.

%%%%%%%%%%%%%%%%%%%%%%%%%%%%%%%%%%%%%%%%%%%%%%%%%%%%%%%%%%%%%%%%%%%%%%%%%%%%%%%%%%%%%
%%%%%%%%%%%%%%%%%%%%%%%%%%%%%%%%%%%%%%%%%%%%%%%%%%%%%%%%%%%%%%%%%%%%%%%%%%%%%%%%%%%%%
%%
%% REFERENCES
%%
%%%%%%%%%%%%%%%%%%%%%%%%%%%%%%%%%%%%%%%%%%%%%%%%%%%%%%%%%%%%%%%%%%%%%%%%%%%%%%%%%%%%%
%%%%%%%%%%%%%%%%%%%%%%%%%%%%%%%%%%%%%%%%%%%%%%%%%%%%%%%%%%%%%%%%%%%%%%%%%%%%%%%%%%%%%

\label{lastpage}

\end{document}